\documentclass[onecolumn, draftcls, 12pt, romanappendices]{IEEEtran}
\usepackage[noadjust]{cite}
\usepackage{graphicx, array, amssymb, psfrag, graphics, setspace, amsmath, booktabs}
\newcommand{\comdots}{, \, \dots \hspace{-1pt} ,}
\title{Equivalent Codes, Optimality, and Performance Analysis of
OSTBC: Textbook Study}
\author{Alex~E.~Geyer, Reza~Nikjah, Sergiy~A.~Vorobyov, and 
Norman~C.~Beaulieu} 

\begin{document}

\maketitle

\begin{abstract}
An equivalent model for a multi-input, multi-output (MIMO)
communication system with orthogonal space-time block codes
(OSTBCs) is proposed based on a newly revealed connection between
OSTBCs and Euclidean codes. Examples of distance spectra, signal
constellations, and signal coordinate diagrams of Euclidean codes
equivalent to simplest OSTBCs are given. A new asymptotic upper
bound for the symbol error rate (SER) of OSTBCs, based on the
distance spectra of the introduced equivalent Euclidean codes, is
derived and new general design criteria for signal constellations
of the optimal OSTBC are proposed. Some bounds relating distance
properties, dimensionality, and cardinality of OSTBCs with
constituent signals of equal energy are given, and new optimal
signal constellations with cardinalities $M = 8$ and $M = 16$ for
Alamouti's code are designed. Using the new model for MIMO
communication systems with OSTBCs, a general methodology for
performance analysis of OSTBCs is developed. As an example of the
application of this methodology, an exact evaluation of the SER of
any OSTBC is given. Namely, a new expression for the SER of
Alamouti's OSTBC with binary phase shift keying (BPSK) signals is
derived.
\end{abstract}

\setcounter{page}{0} \thispagestyle{empty}
\newpage

\section{Introduction}
The simplicity of mathematical description, low complexity of
maximum likelihood (ML) decoding, and unique properties allowing
for noncoherent detection make orthogonal space-time block codes
(OSTBCs) \cite{AlamoutiSM98_JSAC, TarokhV99_TInfo, TarokhV00_JSAC,
MaWK07_TSP} the most attractive and well studied class of
space-time codes. Any OSTBC can be described mathematically by its
corresponding code matrix and a constituent signal
constellation.\footnote{Hereafter, the {\it (OSTBC) signal
constellation} refers to the set of all realizable samples of the
OSTBC matrix, each transmitted in a number of consecutive time
slots, and the {\it constituent signal constellation} refers to
the set of signals that constitutes the components of the OSTBC
matrix. The cardinality of the former constellation is denoted
$M$, and of the latter, is denoted $L$, as defined in Section
\ref{sysmodel}.} Although the code matrices of OSTBCs have been
tabulated for many important cases \cite{AlamoutiSM98_JSAC,
TarokhV99_TInfo, TirkkonenO02_TInfo, SuW03_TInfo, LiangXB03_TInfo,
LiangXB03_ComLetters, SuW03_WPC, SuW04_ComLetters,
GanesanG01_TInfo} using the theory of complex orthogonal design
\cite{TarokhV99_TInfo}, results on the optimal constituent signal
constellations of OSTBCs are extremely limited. In fact, almost
all the investigations of OSTBCs are based on a restricted group
of constituent signals which belong to a class of the
constellations with independent signals. It is, however, unknown
and questionable whether such signal constellations are actually
optimal. Moreover, no general results for guaranteeing the
optimality of OSTBCs are available. It has been stressed, for
example, in \cite{Gharavi-AlkhansariM05_TSP} that general design
criteria for optimal OSTBCs are unknown. Even the problem of
finding constellations optimal in the sense of minimizing an
average error probability of ML decoding on Rayleigh flat fading
channels is an open problem of great interest for multi-input,
multi-output (MIMO) communication systems. The latter problem has
been investigated in \cite{Gharavi-AlkhansariM05_TSP} for the
smallest possible constellations (up to $M = 5$, where $M$ is the
cardinality of an OSTBC signal constellation). Particularly,
OSTBCs that are optimal in the sense of minimizing the symbol
error rate (SER) of ML decoding have been designed only for
constellations with $M = 2 \sim 5$. In these cases, the SER
minimization is equivalent to the minimization of the average
error probability of ML decoding. Despite the aforementioned
limitation of the results in \cite{Gharavi-AlkhansariM05_TSP},
this work is, to the best of the authors' knowledge, the only
source of information available on the design of optimal OSTBCs.

Similarly, although the distance properties of OSTBCs have been
investigated in some previous research works
\cite{ShulzeH03_ComLetters, LarssonEG03,
Gharavi-AlkhansariM05_TInfo}, the distance properties of OSTBC
signal constellations have not attracted any attention. Indeed,
existing results on the distance properties of OSTBCs aim at
verifying the resilience properties of OSTBCs, where a
multidimensional constellation is said to be resilient in flat
fading if it retains its shape when its points are subject to the
multiplicative distortion associated with fading coefficients
\cite{IonescuDM07_TInfo}. However, it is specifically the full
understanding of the distance properties of OSTBC signal
constellations that can enable formulating requirements or design
criteria for OSTBC signal constellations.

In this paper,\footnote{Some initial results have been reported in
\cite{GeyerAE11_Asilomar}.} the aforementioned distance properties
of OSTBCs with arbitrary signal constellations are analyzed. Based
on the analysis, a new equivalent model for a communication system
with orthogonal space-time block coding is proposed. The model is
based on a connection found between the distance properties of
OSTBCs and the distance properties of Euclidean codes, which
allows viewing certain Euclidean codes as equivalent codes to
OSTBCs. This connection brings important insights into OSTBCs
since Euclidean codes fall under the classic theory of error
correcting codes \cite{ShulzeH03_ComLetters} and, thus, the OSTBCs
can now be viewed as a part of the classic theory. Particularly,
the connection between OSTBCs and Euclidean codes enables one to
formulate a new general criterion for designing optimal OSTBCs
with arbitrary constituent signals for the case of large
signal-to-noise ratio (SNR). Indeed, OSTBCs can be viewed as a
subclass of error correcting codes having a specific design
criterion that enables searching for new existence conditions for
optimal OSTBC signal constellations with constant envelope
constituent signals. Such conditions are based on a connection
between the optimal OSTBC signal constellations with equal
energies and a class of spherical codes \cite{EricsonT01}.  For
example, we derive two new optimal biorthogonal signal
constellations with cardinalities $M = 8$ and $M = 16$ for the
Alamouti OSTBC with constant energy signals.

The model introduced for the OSTBC MIMO system can be used for
performance analysis of OSTBCs and enables one to develop a new
performance analysis methodology. Existing results on OSTBC
performance analysis (see \cite{BauchG01_AnnTelecom, GaoC02_VTC,
Gharavi-AlkhansariM04_TCom, ZhangH05_TWireless, LuH03_TInfo,
BehnamfarF05_TCom, SandhuS01_ICASSP, BrehlerM01_TInfo} and the
references therein) aim at deriving exact solutions only for the
SER of the constituent signals of the OSTBC, and there are no
results on the exact solution for the true SER of any OSTBC. As an
example of applying our methodology, we derive a closed-form
solution for the SER of the Alamouti OSTBC with constituent binary
phase shift keying (BPSK) signals. To the best of the authors'
knowledge, this is a new expression for the SER of the Alamouti
OSTBC with constituent BPSK signals. Moreover, this result is the
only exact expression available for the SER of any OSTBC.

The remainder of this paper is organized as follows. In
Section~II, the distance properties of OSTBC signal constellations
are analyzed and a new equivalent model for a MIMO communication
system with orthogonal space-time block coding on a quasistatic
fading channel is given. Some examples of signal coordinate
diagrams and distance spectra of some simplest OSTBCs are also
given. A new union bound on the SER of an OSTBC based on the
distance properties of equivalent Euclidean codes is derived in
Section III. Using this bound, a new general design criterion for
optimal OSTBC constellations and a new design criterion for
optimal constant envelope OSTBC signals as well as some existence
conditions are formulated. As an example of applying this new
design criterion, two new optimal biorthogonal constellations for
the Alamouti OSTBC with $M = 8$ and $M = 16$ are designed. In
Section IV, a new general OSTBC performance analysis methodology
based on the equivalent model for MIMO systems introduced in
Section II is described. A new closed-form solution for the SER of
the Alamouti OSTBC with constituent BPSK signals is also derived.
Section V presents some numerical examples and is followed by some
conclusions in Section VI.

\section{New System Model and Equivalent Codes}\label{sysmodel}
In this section, a new model for a communication system with an
OSTBC having an arbitrary signal constellation is introduced.
Toward this end, a modified description of an OSTBC with an
arbitrary signal constellation, its distance properties and its
connections to the class of Euclidean codes are of interest. Based
on the new model, examples of signal constellations and signal
coordinate diagrams of Euclidean codes equivalent to the simplest
OSTBC are developed.

\subsection{OSTBCs With Arbitrary Constituent Signals} \label{sysdef}
An OSTBC can be defined by a generalized complex orthogonal design
\cite{TarokhV99_TInfo}, i.e., by an $N_T \times N_T$ code matrix
${\boldsymbol G}_u$ with orthogonal columns. The entries $g_{i,
j}^u$ ($i, j = 1 \comdots N_T$) of the code matrix ${\boldsymbol
G}_u$ are the elements $s_{t, u}$ ($t = 1 \comdots N_t; \; u = 0
\comdots M - 1$) of the codewords (signal constellations)
\begin{equation}\label{equation1}
    {\boldsymbol s}_u = \left[s_{1, u}, s_{2, u} \comdots
    s_{N_T, u}\right]^T, \qquad u = 0 \comdots M - 1
\end{equation}
as well as the complex conjugates $s_{t, u}^\ast$ ($t = 1 \comdots
N_T; \; u = 0 \comdots M - 1$), linear combinations of $s_{t, u}$
($t = 1 \comdots N_T; \; u = 0 \comdots M - 1$) and $s_{t,
u}^\ast$ ($t = 1 \comdots N_T; \; u = 0 \comdots M - 1$), or
zeros. Here $[\cdot]^T$ is the matrix transpose and $M$ is the
cardinality of the OSTBC signal constellation. The codewords
\eqref{equation1} belong to a block code with $J$ constituent 1-
or 2-dimensional (1- or 2-D) signals $s_{t, u}$ ($t = 1 \comdots
J; \; u = 0 \comdots M - 1$) and $N_T - J$ zero signals $s_{t, u}
= 0$ ($t = J + 1 \comdots N_T; \; u = 0 \comdots M - 1$) with $J
\leq N_T$ denoting the number of information bearing constituent
signals of the OSTBC. Since the multidimensional signal
constellations \eqref{equation1} belong to a block code
constructed of modulated symbols from its alphabet, such a code
corresponds to the so-called Euclidean code known from the classic
theory of error correcting coding. This connection helps to define
a complex structure of $M$-ary constellations belonging to OSTBCs.

{\bf Definition 1} \cite{ZegerK94_TInfo}{\bf:} The {\it Euclidean
code} is a finite set of $M$ points (codewords) in $n$-D Euclidean
space $R^n$.

The constituent signals $s_{t, u}$ ($t = 1 \comdots N_T; \; u = 0
\comdots M - 1$) of a \textit{canonical} OSTBC
\cite{AlamoutiSM98_JSAC, TarokhV99_TInfo} use the same, typically
$L$-PSK or $L$-QAM, modulation with $L$ being the cardinality of
the constellation. Note, however, that in the general case, the
constituent symbols of the Euclidean code $s_{t, u}$ ($t = 1
\comdots N_T; \; u = 0 \comdots M - 1$) can have arbitrary signal
constellations including correlated constellations.

Assuming a flat fading channel, the signal received by the $j$th
receiving antenna ($j = 1$ $\comdots$ $N_R$) of the OSTBC MIMO
system can be expressed as
\begin{equation}\label{eq3}
    {\boldsymbol r}_j = {\boldsymbol G}_u \, {\boldsymbol h}_j +
    {\boldsymbol n}_j, \quad j = 1 \comdots N_R
\end{equation}
where ${\boldsymbol h}_j =\left[h_{1, j} \comdots h_{N_T, j}
\right]^T$ is the $N_T \times 1$ vector of fading channel
coefficients, which are assumed to be independent, identically
distributed zero-mean complex Gaussian variables with variance
$\rho / 2$ per dimension and are assumed constant over $N_T$ (or
some multiple of $N_T$) time periods; ${\boldsymbol n}_j =
\left[n_{1, j} \comdots n_{N_T, j}\right]^T$ is an $N_R \times 1$
vector of complex Gaussian additive noises consisting of
independent samples of zero-mean complex Gaussian random variables
each of variance $N_0 / 2$ per dimension; and ${\boldsymbol r}_j
=\left[r_{1, j} \comdots r_{N_T, j}\right]^T$ is the $N_T \times
1$ received signal vector at the $j$th receiving antenna. It can
be observed from \eqref{eq3} that if the OSTBC codewords occur
with equal probability, the average received SNR per antenna is
given by
\begin{equation}\label{avgsnr}
    \mathrm{SNR_R} \triangleq \frac{\rho \sum_{u = 0}^{M - 1}
    \|{\boldsymbol G}_u\|_F^2}{M N_T N_0}
\end{equation}
where $\|\cdot\|_F$ is the Frobenius norm of a matrix
\cite{HornRA90}.

\subsection{Distance Properties and Equivalent Model}
According to the classic approach of analyzing any type of
modulation or coding schemes, the distance properties (signal
coordinate diagrams) of the signals under consideration should be
first studied. To study the distance properties, we assume the 
noise-free case. Then the Euclidean distance between received 
noise-free codeword vectors ${\boldsymbol G}_u \, {\boldsymbol h}_j $ 
and ${\boldsymbol G}_t \, {\boldsymbol h}_j$ ($u \neq t$) of an OSTBC,
denoted as $d_{u, t, \rm{OSTBC}}^j$, can be expressed in terms of
the distance for the equivalent Euclidean code $d_{u, t, {\rm EC}}
\triangleq \left\|{\boldsymbol s}_u - {\boldsymbol s}_t\right\|$
as
\begin{equation}\label{eq7}
    d_{u, t, \rm{OSTBC}}^j = \|{\boldsymbol h}_j\| \, d_{u, t,
    {\rm EC}}, \quad j = 1 \comdots N_R; \; u, t = 0 \comdots M
    - 1; \; u \neq t
\end{equation}
where $\|\cdot\|$ denotes the Euclidean norm of a vector.

Indeed, the distances $d_{u, t, \rm{OSTBC}}^j$ are defined as
\begin{equation}\label{eq8}
    d_{u, t, \rm{OSTBC}}^j \triangleq \left\|{\boldsymbol G}_u
    \, {\boldsymbol h}_j - {\boldsymbol G}_t \, {\boldsymbol h}_j
    \right\| = \left\|\left({\boldsymbol G}_u - {\boldsymbol G}_t
    \right) {\boldsymbol h}_j\right\|, \quad j = 1 \comdots N_R;
    \; u, t = 0 \comdots M - 1; \; u \neq t.
\end{equation}
Using the OSTBC orthogonality property, and the property that
\cite[p.~120]{LarssonEG03}
\begin{equation}\label{eq9}
    ({\boldsymbol G}_u - {\boldsymbol G}_t)^H ({\boldsymbol G}_u
    - {\boldsymbol G}_t) = \|{\boldsymbol s}_u - {\boldsymbol s}_t
    \|^2 {\boldsymbol I}_{N_T}
\end{equation}
where $(\cdot)^H$ denotes the Hermitian transpose, the distances
\eqref{eq8} can be written as in \eqref{eq7} by also noting that
$\|{\boldsymbol s}_u - {\boldsymbol s}_t \| \triangleq d_{u, t,
{\rm EC}}$. Here ${\boldsymbol I}_{N_T}$ is the $N_T \times N_T$
identity matrix. Moreover, using the OSTBC orthogonality property,
the norms of the received noise-free codeword vectors
${\boldsymbol G}_u \, {\boldsymbol h}_j$ ($u = 0 \comdots M - 1;
\; j = 1 \comdots N_R$) can be computed as
\begin{equation}\label{eq11}
    \left\|{\boldsymbol G}_u {\boldsymbol h}_j\right\| =
    \sqrt{{\boldsymbol h}_j^H {\boldsymbol G}_u^H {\boldsymbol G}_u
    {\boldsymbol h}_j} = \left\|{\boldsymbol s}_u\right\| \cdot
    \|{\boldsymbol h}_j\|, \quad u = 0 \comdots M - 1, \quad j
    = 1 \comdots N_R.
\end{equation}

On the other hand, let ${\boldsymbol \Phi}_j$ be an $N_T \times
N_T$ arbitrary unitary matrix, i.e. ${\boldsymbol \Phi}_j^H
{\boldsymbol \Phi}_j = {\boldsymbol \Phi}_j {\boldsymbol \Phi}_j^H
= {\boldsymbol I}_{N_T}$. Now consider the new constellation
$\|{\boldsymbol h}_j\| {\boldsymbol \Phi}_j {\boldsymbol s}_u$, $u
= 0 \comdots M - 1$. It can be observed that this constellation
has exactly the same Euclidean distance properties \eqref{eq7} and
\eqref{eq11} of the fading-inflicted constellation ${\boldsymbol
G}_u {\boldsymbol h}_j$, $u = 0 \comdots M - 1$.

The following theorem can now be formulated based on the distance
properties cited above.

\begin{figure}
    \psfrag{Data}{Data}
    \psfrag{Encoder}[][]{\parbox{70pt}{\centering Encoder of
    Euclidean code ($\boldsymbol{s}_u, u = 0 \comdots M - 1$)}}
    \psfrag{C1}[][]{$c_1 {\boldsymbol \Phi}_1$}
    \psfrag{C2}[][]{$c_2 {\boldsymbol \Phi}_2$}
    \psfrag{CNR}[][]{$c_{N_R} {\boldsymbol \Phi}_{N_R}$}
    \psfrag{Cj=sqrt}{\hspace{-20pt}\parbox{120pt}{$c_j =
    \sqrt{\sum_{i = 1}^{N_T} |h_{i, j}|^2}$\\[10pt]${\boldsymbol
    \Phi}_j^H {\boldsymbol \Phi}_j = {\boldsymbol \Phi}_j
    {\boldsymbol \Phi}_j^H = I_{N_T}$}}
    \psfrag{n1}[][]{${\boldsymbol n}_1$}
    \psfrag{n2}[][]{${\boldsymbol n}_2$}
    \psfrag{nNR}[][]{${\boldsymbol n}_{N_R}$}
    \psfrag{ML}[][]{ML decoder}
    \psfrag{Decision}{Decision}
    \psfrag{AWGN}{AWGN}
    \psfrag{su}{$\boldsymbol{s}_u$}
    \vspace{100pt}
    \centering
  \includegraphics[width=380pt]{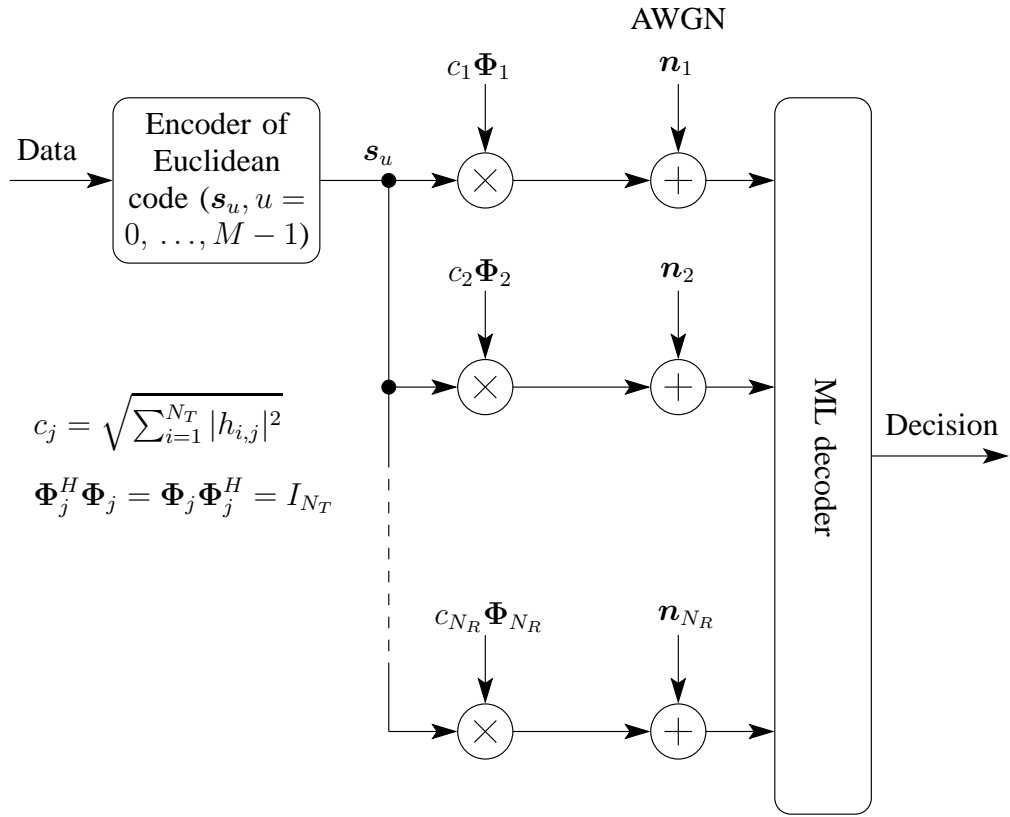}
    \caption{Equivalent model of a communication system with
    OSTBC and ML decoding.}
    \vspace{100pt}
    \label{fig_EquiModel}
\end{figure}

{\bf Theorem 1:} {\it A communication system with $N_T$
transmitting and $N_R $ receiving antennas, OSTBC, and maximum
likelihood decoding of received signals is equivalent to the
system given in Fig. \ref{fig_EquiModel} for Gaussian noise
channels.}

{\bf Proof:} The statement of the theorem directly follows from
the properties \eqref{eq7} and \eqref{eq11}, and the fact that two
codes (signals) with the same Euclidean distance properties
provide the same performance with maximum likelihood decoding in
the Gaussian noise channel \cite{LinS04}.\footnote{The claim of
the theorem may not hold in channels with other types of noise,
such as Laplacian noise \cite{ShaoH12_TWireless}.} $\hfill\Box$

Note that the matrix ${\boldsymbol \Phi}_j$ is, in fact, a
rotation matrix in $N_T$ dimensions. What can be seen from Theorem
1 and Fig. \ref{fig_EquiModel} is that the OSTBC effectively
transforms the fading MIMO channel into an equivalent coded
single-input, multi-output (SIMO) channel with the corresponding
fading coefficients $c_j \triangleq \|{\boldsymbol h}_j \|$ ($j =
1 \comdots N_R$) (and, therefore, the probability of deep fading
in the channels of the equivalent SIMO system is lower than that
in the actual channels of the MIMO system). This SIMO channel is
invariant to phase rotation in the sense that different arbitrary
rotation matrices ${\boldsymbol \Phi}_j$ ($j = 1 \comdots N_R$)
give rise to the same ML performance.

Note also that the resulting SNR in the equivalent model, Fig.
\ref{fig_EquiModel}, is not always equal to the original average
SNR \eqref{avgsnr}, if we exclude the zeros, i.e.
non-information-bearing components, of $\boldsymbol{s}_u$ in the
equivalent Euclidean code of the equivalent model (see Fig.
\ref{fig_EquiModel}). In fact in this case, recalling the
definition of $J$ in Section \ref{sysdef}, we can show by energy
conservation that the average SNR in the equivalent model is $N_T
/ J$ times the SNR \eqref{avgsnr}.

It is also worth stressing that the system model in Fig.
\ref{fig_EquiModel} is a special case of a receiver diversity
system \cite{ProakisJG08, BauchG01_AnnTelecom}. However, it is
also important to note that the proposed coded SIMO model is
fundamentally different from the well-known single-input,
single-output (SISO) model of \cite{BauchG01_AnnTelecom} because
(i) it represents the actual multidimensional structure of $M$-ary
OSTBCs; (ii) it is applicable to arbitrary constituent signal
constellations of an OSTBC; and (iii) it allows using any existing
receiver diversity schemes.

It is worth highlighting as well that as follows from
Fig.~\ref{fig_EquiModel}, Euclidean codes are equivalent to OSTBCs
in the sense that the parameters of the Euclidean codes are the
only available optimization parameters for optimizing the MIMO
communication system with OSTBC (see Fig. \ref{fig_EquiModel}).
Therefore, the OSTBC design is equivalent to Euclidean code design
from the system point of view.

Most of the known OSTBCs belong to a subclass of canonical OSTBCs,
i.e., to the class of codes based on signal constellations with
uncorrelated constituent signals. However, this condition is
extremely restrictive for designing good signal constellations of
OSTBCs, while it is clearly not necessary or particularly
appealing from a practical (decoder complexity) viewpoint. As a
result, OSTBCs with different or correlated constellations for the
constituent signals and their properties are essentially
overlooked and have not been studied. Therefore, we aim at
correcting this deficiency in the existing literature by providing
a detailed analysis and design criteria for such codes in
Section~III.

Toward this end, we first explicitly connect the terminology used
to describe the signal constellations of OSTBC with the
terminology commonly used for describing error correction codes.
Particularly, we define the Euclidean code equivalent to a given
OSTBC as follows.

{\bf Definition 2:} The Euclidean code $\tilde{\boldsymbol s}_u =
[\tilde{s}_{1, u} \comdots \tilde{s}_{K, u}]$ is called equivalent
to an OSTBC with ``proxy'' Euclidean code ${\boldsymbol s}_u =
[s_{1, u} \comdots s_{N_T, u}]$ if the Euclidean distance between
two arbitrary codewords ${\boldsymbol s}_u $ and ${\boldsymbol
s}_t$ of the OSTBC coincide with the distance between
$\tilde{\boldsymbol s}_u $ and $\tilde{\boldsymbol s}_t$ for all
$u, t \in \{0 \comdots M - 1\}$. In other words, the distances
between two codeword vectors ${\boldsymbol G}_u \, {\boldsymbol
h}_j$ and ${\boldsymbol G}_t \, {\boldsymbol h}_j$ of the OSTBC
coincide with the distances between $\tilde{\boldsymbol s}_u $ and
$\tilde{\boldsymbol s}_t$ for all $u, t \in \{0 \comdots M - 1\}$,
that is, $d_{u, t, {\rm EC}}$ and $d_{u, t, \rm{OSTBC}}^j$ satisfy
\eqref{eq7} for all $j = 1 \comdots N_R$.

It follows from Definition 2 that if an OSTBC code matrix does not
contain any zeros, then the equivalent Euclidean code coincides
with the ``proxy'' Euclidean code of the OSTBC. The salient
example is Alamouti's code. However, if the code matrix contains
zeros, the dimensionality of the equivalent Euclidean code is
smaller. For example, the equivalent Euclidean code to the rate
$3/4$ OSTBC with the following code matrix \cite{GanesanG01_TInfo}
\begin{equation}\label{eq13}
    {\boldsymbol G}_u = \left[
    \begin{array}{*{20}c}
        s_{1, u} \hfill & 0 \hfill & s_{2, u} \hfill & -s_{3, u}
        \hfill \\
        0 \hfill & s_{1, u} \hfill & s_{3, u}^\ast \hfill & s_{2,
        u}^\ast \hfill \\
        -s_{2, u}^\ast \hfill & -s_{3, u} \hfill & s_{1, u}^\ast
        \hfill & 0 \hfill \\
        s_{3, u}^\ast \hfill & -s_{2, u} \hfill & 0 \hfill &
        s_{1, u}^\ast \hfill \\
    \end{array}
    \right], \quad u = 0 \comdots M - 1
\end{equation}
and the Euclidean code ${\boldsymbol s}_u = [s_{1, u}, s_{2, u},
s_{3, u}, 0]$ is $\tilde{\boldsymbol s}_u = [\tilde{s}_{1, u},
\tilde{s}_{2, u}, \tilde{s}_{3, u}]$.

\subsection{Examples of Euclidean Codes Equivalent to Some
Simplest OSTBCs}
Traditionally, only distance properties of OSTBC matrices have
been investigated, and this was believed to be sufficient (see,
for example, \cite{LarssonEG03}). Geometrical properties of OSTBCs
are discussed in several papers such as
\cite{ShulzeH03_ComLetters, LarssonEG03,
Gharavi-AlkhansariM05_TInfo}. However, to the best of the authors'
knowledge, there is no research which reports the distance
properties of signal coordinate diagrams even for the simplest
OSTBCs. Thus, in this subsection, the distance properties of
Euclidean codes equivalent to some simplest OSTBCs are studied.

The following definitions are needed for further discussion.

{\bf Definition 3:} A distance profile ${\cal D}_u$ of a codeword
${\boldsymbol s}_u$ is a set of Euclidean distances $d_{u, j, {\rm
EC}}$ between the codeword ${\boldsymbol s}_u $ and all other
codewords ${\boldsymbol s}_t$ ($t = 0 \comdots M - 1; \; t \neq
u$).

{\bf Definition 4} \cite{ForneyGD91_TInfo}{\bf:} A code has a
uniform constellation if all its codewords have the same distance
profile. This means that all sets of distances between any
codewords of the code are the same, and therefore, the
corresponding average error probabilities under maximum likelihood
decoding are the same for all codewords.

{\bf Definition 5:} If a code has a uniform constellation, the
corresponding distance profile is called a distance spectrum.

Let the normalized Euclidean distance be defined as
\begin{equation}\label{eq14}
    \tilde{d}_{u, t, {\rm EC}} \triangleq \frac{d_{u, t,
    {\rm EC}}}{\sqrt{\bar{E}_{\rm EC}}}
\end{equation}
where $\bar{E}_{\rm EC}$ is defined as the average energy of a
codeword of the Euclidean code. The normalized distance spectra of
two Euclidean codes equivalent to the simplest Alamouti OSTBC with
constituent BPSK and quadrature PSK (QPSK) signals are given in
Tables \ref{tab_alambpsk} and \ref{tab_alamqpsk}, respectively.
These spectra are calculated according to \eqref{eq14}. In Table
\ref{tab_stbcbpsk} the normalized distance spectrum of the rate
3/4 OSTBC with the code matrix \eqref{eq13} and the equivalent
Euclidean code with constituent BPSK signals is also given.
Moreover, the average energies of the codes are $\bar{E}_{\rm EC}
= 2 E$ for the Alamouti OSTBC with constituent BPSK and QPSK
signals and $\bar{E}_{\rm EC} = 3 E$ for the rate 3/4 OSTBC
\eqref{eq13} with constituent BPSK signals, where $E$ is the
energy of a constituent signal of the code. All these codes have
uniform signal constellations in the equivalent Euclidean codes.

\begin{table}
\caption{Distance Spectrum of the Euclidean Code Equivalent to
Alamouti's Code\protect\\With Constituent BPSK Signals}
\vspace{-20pt}
\begin{center}
    \begin{tabular}{lcc}
        \toprule\addlinespace[1pt]
        \rule{0pt}{13pt} Normalized Euclidean distance
        & ~~$\sqrt{2}$~~ & ~~$2$~~ \\\addlinespace[1pt]
        \hline
        \rule{0pt}{13pt} Number of codewords & ~~$2$~~ & ~~$1$~~ \\
        \bottomrule
    \end{tabular}
\end{center}
\label{tab_alambpsk}
\end{table}

\begin{table}
\caption{Distance Spectrum of the Euclidean Code Equivalent to
Alamouti's Code\protect\\With Constituent QPSK Signals}
\vspace{-20pt}
\begin{center}
    \begin{tabular}{lcccc}
        \toprule\addlinespace[1pt]
        \rule{0pt}{13pt} Normalized Euclidean distance
        & ~~$1$~~ & ~~$\sqrt{2}$~~ & ~~$\sqrt{3}$~~ & ~~$2$~~ \\\addlinespace[1pt]
        \hline
        \rule{0pt}{13pt} Number of codewords & ~~$4$~~ & ~~$6$~~ & ~~$4$~~ & ~~$1$~~ \\
        \bottomrule
    \end{tabular}
\end{center}
\label{tab_alamqpsk}
\end{table}

\begin{table}
\caption{Distance Spectrum of the Euclidean Code Equivalent to the
Rate $3 / 4$ OSTBC \eqref{eq13}\protect\\With Constituent BPSK
Signals} \vspace{-20pt}
\begin{center}
    \begin{tabular}{lccc}
        \toprule\addlinespace[1pt]
        \rule{0pt}{17pt} Normalized Euclidean distance
        & ~~$2 \sqrt{\frac{1}{3}}$~~ & ~~$2 \sqrt{\frac{2}{3}}$~~ & ~~$2$~~
        \\\addlinespace[4pt]
        \hline
        \rule{0pt}{13pt} Number of codewords & ~~$3$~~ & ~~$3$~~ & ~~$1$~~ \\
        \bottomrule
    \end{tabular}
\end{center}
\label{tab_stbcbpsk}
\end{table}

\begin{figure}
    \psfrag{S0, S1}{$\boldsymbol{s}_0, \boldsymbol{s}_1$}
    \psfrag{S0, S2}{$\boldsymbol{s}_0, \boldsymbol{s}_2$}
    \psfrag{S2, S3}{$\boldsymbol{s}_2, \boldsymbol{s}_3$}
    \psfrag{S1, S3}{$\boldsymbol{s}_1, \boldsymbol{s}_3$}
    \psfrag{S1u}{$s_{1, u}$}
    \psfrag{S2u}{$s_{2, u}$}
    \psfrag{E}[l][l]{$\sqrt{E} \left\{\rule{0pt}{32pt}\right.$}
    \centering
    \includegraphics{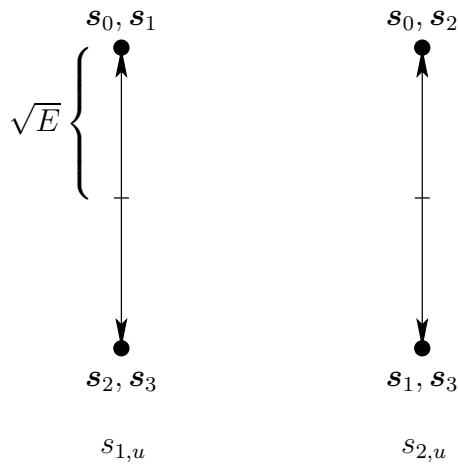}
    \vspace{-10pt}
    \caption{Signal constellation of the Euclidean code equivalent
    to Alamouti's code with $M = 4$ and constituent BPSK signals.}
    \label{fig_SigAlamBPSK}
\end{figure}

\begin{figure}
    \psfrag{S0}{$\boldsymbol{s}_0$}
    \psfrag{S1}{$\boldsymbol{s}_1$}
    \psfrag{S2}{$\boldsymbol{s}_2$}
    \psfrag{S3}{$\boldsymbol{s}_3$}
    \psfrag{s}[l][l]{$\sqrt{2}$}
    \centering
    \includegraphics{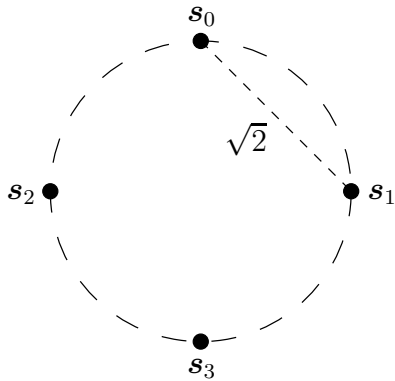}
    \vspace{-10pt}
    \caption{Normalized signal coordinate diagram of the Euclidean
    code equivalent to Alamouti's code with $M = 4$ and constituent
    BPSK signals.}
    \label{fig_VecAlamBPSK}
\end{figure}

\begin{figure}
    \psfrag{S0, S1, S2, S3}[][]{$\boldsymbol{s}_0, \boldsymbol{s}_1, \boldsymbol{s}_2, \boldsymbol{s}_3$}
    \psfrag{S0, S4, S8, S12}[][]{$\boldsymbol{s}_0, \boldsymbol{s}_4, \boldsymbol{s}_8, \boldsymbol{s}_{12}$}
    \psfrag{S8, S9, S10, S11}[l][l]{\parbox{40pt}{\centering$\boldsymbol{s}_8$, $\boldsymbol{s}_9$, $\boldsymbol{s}_{10}$, $\boldsymbol{s}_{11}$}}
    \psfrag{S1, S5, S9, S13}[l][]{\parbox{40pt}{\centering$\boldsymbol{s}_1$, $\boldsymbol{s}_5$, $\boldsymbol{s}_9$, $\boldsymbol{s}_{13}$}}
    \psfrag{S12, S13, S14, S15}[][]{$\boldsymbol{s}_{12}, \boldsymbol{s}_{13}, \boldsymbol{s}_{14}, \boldsymbol{s}_{15}$}
    \psfrag{S3, S7, S11, S15}[][]{$\boldsymbol{s}_3, \boldsymbol{s}_7, \boldsymbol{s}_{11}, \boldsymbol{s}_{15}$}
    \psfrag{S4, S5, S6, S7}[l][l]{\parbox{40pt}{\centering$\boldsymbol{s}_4$, $\boldsymbol{s}_5$,\\$\boldsymbol{s}_6$, $\boldsymbol{s}_7$}}
    \psfrag{S2, S6, S10, S14}[l][l]{\parbox{40pt}{\centering$\boldsymbol{s}_2$, $\boldsymbol{s}_6$,\\$\boldsymbol{s}_{10}$, $\boldsymbol{s}_{14}$}}
    \psfrag{S1u}{$s_{1, u}$}
    \psfrag{S2u}{$s_{2, u}$}
    \centering
    \includegraphics{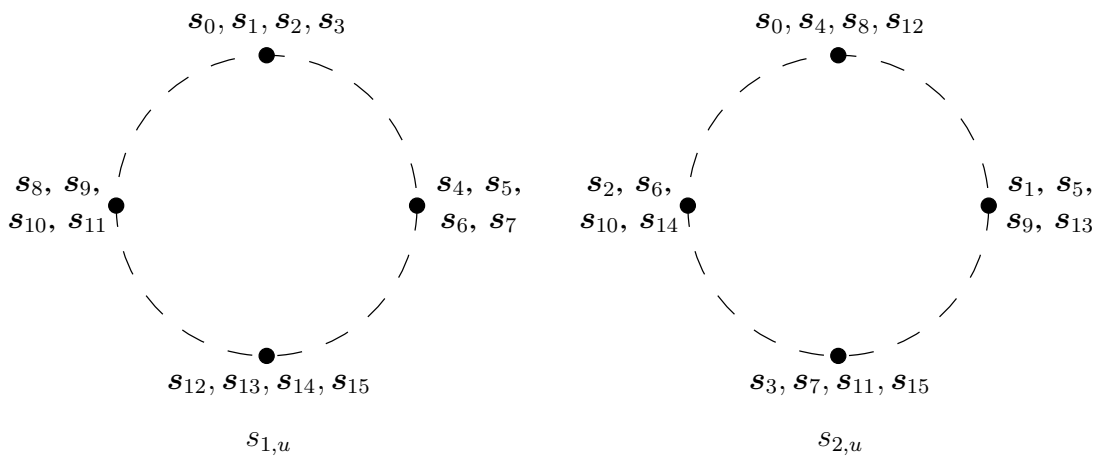}
    \vspace{-5pt}
    \caption{Signal constellation of the Euclidean code equivalent to Alamouti's
    code with $M = 16$ and constituent QPSK signals with Grey mapping.}
    \label{fig_SigAlamQPSK}
\end{figure}

\begin{figure}
    \psfrag{S0}{\large$\boldsymbol{s}_0$}
    \psfrag{S1}{\large$\boldsymbol{s}_1$}
    \psfrag{S2}{\large$\boldsymbol{s}_2$}
    \psfrag{S3}{\large$\boldsymbol{s}_3$}
    \psfrag{S4}{\large$\boldsymbol{s}_4$}
    \psfrag{S5}{\large$\boldsymbol{s}_5$}
    \psfrag{S6}{\large$\boldsymbol{s}_6$}
    \psfrag{S7}{\large$\boldsymbol{s}_7$}
    \psfrag{S8}{\large$\boldsymbol{s}_8$}
    \psfrag{S9}{\large$\boldsymbol{s}_9$}
    \psfrag{S10}{\large$\boldsymbol{s}_{10}$}
    \psfrag{S11}{\large$\boldsymbol{s}_{11}$}
    \psfrag{S12}{\large$\boldsymbol{s}_{12}$}
    \psfrag{S13}{\large$\boldsymbol{s}_{13}$}
    \psfrag{S14}{\large$\boldsymbol{s}_{14}$}
    \psfrag{S15}{\large$\boldsymbol{s}_{15}$}
    \psfrag{1}{\large$1$}
    \centering
    \includegraphics[scale=.9]{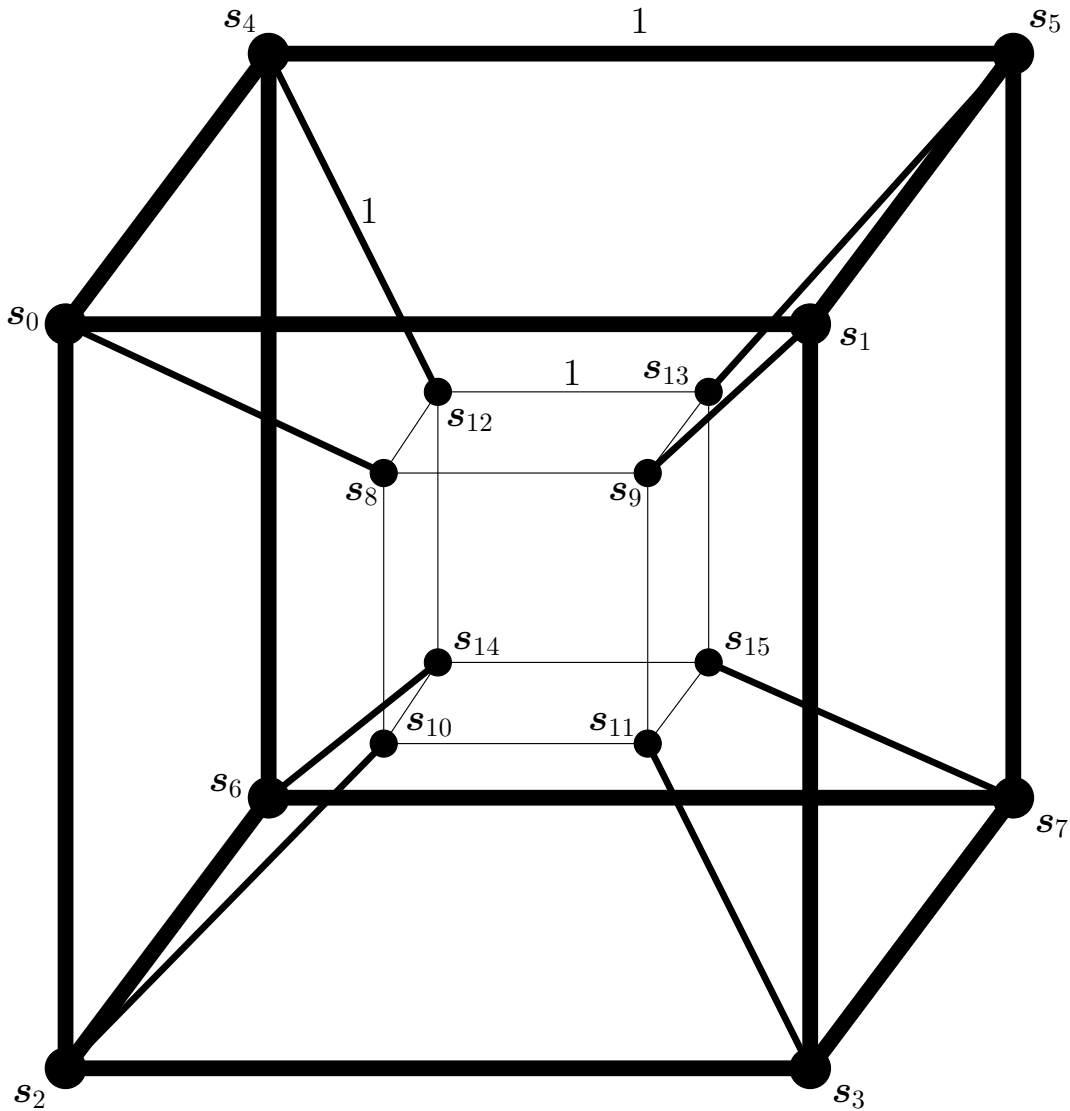}
    \caption{The Schlegel diagram of the tesseract (hypercube) in the
    4-D space [after http://en.wikipedia.org/wiki/Tesseract] showing a
    geometrical representation of the Euclidean code equivalent to
    Alamouti's code with $M = 16$ and constituent QPSK signals.}
    \label{fig_VecAlamQPSK}
\end{figure}

\begin{figure}
    \psfrag{S0, S1, S2, S3}{$\boldsymbol{s}_0, \boldsymbol{s}_1, \boldsymbol{s}_2, \boldsymbol{s}_3$}
    \psfrag{S0, S1, S4, S5}{$\boldsymbol{s}_0, \boldsymbol{s}_1, \boldsymbol{s}_4, \boldsymbol{s}_5$}
    \psfrag{S0, S2, S4, S6}{$\boldsymbol{s}_0, \boldsymbol{s}_2, \boldsymbol{s}_4, \boldsymbol{s}_6$}
    \psfrag{S4, S5, S6, S7}{$\boldsymbol{s}_4, \boldsymbol{s}_5, \boldsymbol{s}_6, \boldsymbol{s}_7$}
    \psfrag{S2, S3, S6, S7}{$\boldsymbol{s}_2, \boldsymbol{s}_3, \boldsymbol{s}_6, \boldsymbol{s}_7$}
    \psfrag{S1, S3, S5, S7}{$\boldsymbol{s}_1, \boldsymbol{s}_3, \boldsymbol{s}_5, \boldsymbol{s}_7$}
    \psfrag{S1u}{$s_{1, u}$}
    \psfrag{S2u}{$s_{2, u}$}
    \psfrag{S3u}{$s_{3, u}$}
    \psfrag{E}[l][l]{$\sqrt{E} \left\{\rule{0pt}{32pt}\right.$}
    \centering
    \includegraphics{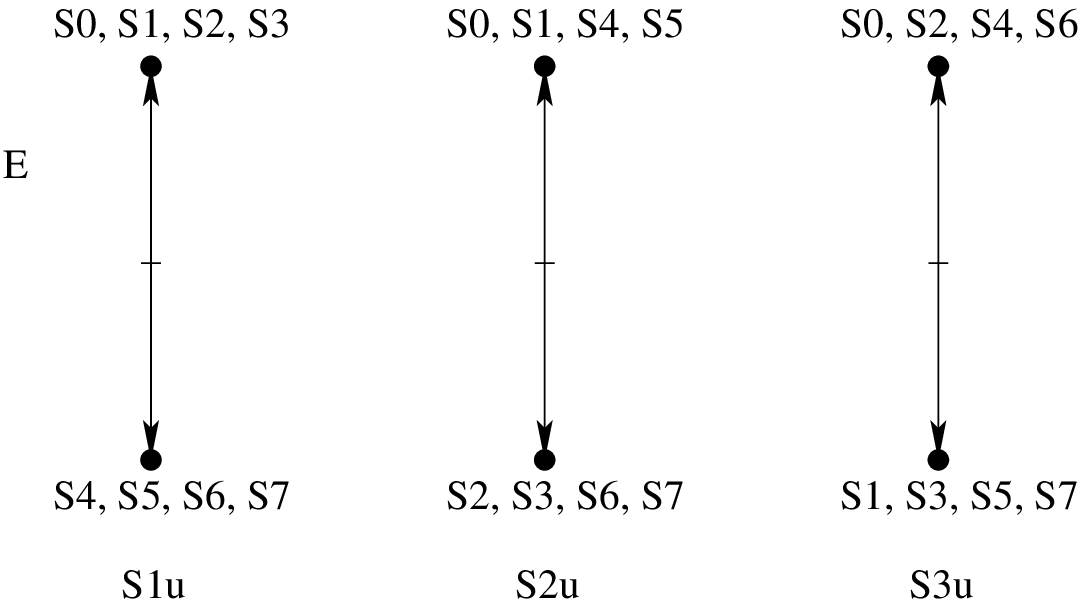}
    \vspace{-5pt}
    \caption{Signal constellation of the Euclidean code with $M = 8$
    equivalent to OSTBC \eqref{eq13} with constituent BPSK signals.}
    \label{fig_SigSTBC_BPSK}
\end{figure}

The corresponding signal constellations and signal coordinate
diagrams (graphical representations) of the Euclidean codes
equivalent to OSTBCs with the spectra given in Tables
\ref{tab_alambpsk}--\ref{tab_stbcbpsk} are illustrated in Figs.
\ref{fig_SigAlamBPSK}--\ref{fig_SigSTBC_BPSK}, where Figs.
\ref{fig_SigAlamBPSK} and \ref{fig_VecAlamBPSK}, correspond to
Table \ref{tab_alambpsk}; Figs. \ref{fig_SigAlamQPSK} and 
\ref{fig_VecAlamQPSK} correspond to Table \ref{tab_alamqpsk}; 
and Fig. \ref{fig_SigSTBC_BPSK} correspond to Table
\ref{tab_stbcbpsk} while the corresponding signal coordinate 
diagram is a simple cube with vertices corresponding to 8 
codewords and edges of length $2/\sqrt{3}$. Note that the 
codes are represented using the notation introduced in
\cite{ZetterbergLH77_TCom} for 4-D group codes. In these figures,
$\boldsymbol{s}_u$ ($u = 0 \comdots M - 1$) is the codeword
transmitted for binary $u$. In Figs. \ref{fig_SigAlamBPSK},
\ref{fig_SigAlamQPSK}, and \ref{fig_SigSTBC_BPSK}, the
constellation points transmitted for $\boldsymbol{s}_u$ are
labeled by $\boldsymbol{s}_u$ itself, for simplicity. Note that in
all figures, a Grey mapping scheme is followed, where the
Euclidean distance between codewords $\boldsymbol{s}_u$ and
$\boldsymbol{s}_t$ is nondecreasing as the Hamming distance
between binary $u$ and binary $t$ increases. For example for $M =
8$, $\boldsymbol{s}_0$ and $\boldsymbol{s}_7$ or
$\boldsymbol{s}_2$ and $\boldsymbol{s}_5$ have the largest
distance. Fig. \ref{fig_VecAlamQPSK} and the cube in the 3-D 
space representing the Euclidean code with $M = 8$ indicated
in Fig. \ref{fig_SigSTBC_BPSK}, equivalent to OSTBC \eqref{eq13}
with constituent BPSK signals use, in fact, the Schlegel diagram 
\cite{CoxeterHSM1883} to provide a geometrical representation of 
the signals and codes.

The tesseract depicted in Fig. \ref{fig_VecAlamQPSK} is an example
of the 4-D Euclidean code (group code) well defined in 4-D
geometry \cite{IngemarssonI89_LNCIS}. Note that the Euclidean
codes given in Figs. \ref{fig_SigAlamBPSK}, \ref{fig_SigAlamQPSK},
and \ref{fig_SigSTBC_BPSK} belong to the class of spherical codes
\cite{EricsonT01} and are also group codes
\cite{IngemarssonI89_LNCIS}.

{\bf Remark 1:} Typically, the canonical OSTBCs are defined in the
literature as OSTBCs with independent information-bearing PSK
signals, as is the case with the examples given above. However, it
should be noted that independent information-bearing signals are
just the $L$-ary auxiliary `components' of a spatial modulator
generating the actual $M$-ary multidimensional signal
constellation of the OSTBC with correlated signals.

\section{Optimality of OSTBCs}\label{optostbc}
A cornerstone of designing `optimal' codes is a proper definition
of the design optimality criterion. Although there is a number of
different definitions of OSTBC optimality (see, for example,
\cite{TarokhV99_TInfo, Gharavi-AlkhansariM05_TSP}), the most
natural one is the following definition which is commonly used for
modulated/coded signals in communication systems.

{\bf Definition 6:} An OSTBC with given constituent signals and
$M$ codewords is called optimal for a given type of channel if it
provides the smallest SER under ML decoding among all OSTBCs with
the same number of codewords, $M$, and arbitrary constituent
signals.

Note that the design criteria for optimal codes are typically
based on connecting the asymptotic SER behavior of a code for a
given channel under a given decoding algorithm, with distance
properties of this code. Thus, in this section, we aim at deriving
a general design criterion for the optimal signal constellations
of OSTBCs by connecting the distance properties of the Euclidean
codes equivalent to OSTBCs to the asymptotic SER performance of
OSTBCs. Then the design criterion for the particular case of large
SNR and a large number of antennas is analyzed and existence
conditions for the optimal signal constellations of OSTBCs with
constituent signals having constant energy are provided,
exploiting the connection of such OSTBCs with the optimal
spherical codes. A new biorthogonal constellation, which is an
example of the optimal signal constellation for the Alamouti
OSTBC, is also given.

\subsection{Union Bound on the SER of OSTBCs for Rayleigh
Fading Channels} The interest in the union bound on the SER of
OSTBCs in Rayleigh fading channels is motivated by the need to
connect the distance spectra of equivalent Euclidean codes with
the asymptotic properties of OSTBCs. As noted in the beginning of
this section, this connection will be used for formulating design
criteria for equivalent Euclidean codes (i.e. constituent
multidimensional signals) of the optimal OSTBC for the Rayleigh
fading channel.

Different upper bounds on the SER of OSTBCs have been previously
derived in, for example, \cite{MaWK07_TSP,
Gharavi-AlkhansariM05_TSP, LuH03_TInfo, BehnamfarF05_TCom,
SandhuS01_ICASSP}. However, one of the most often used upper
bounds on the SER of OSTBCs is a union bound, which can be written
for codes with uniform constellations as
\begin{equation}\label{eq14a}
    {\Pr}_{s, \rm{OSTBC}} \leq \sum_{t = 0, t \neq u}^{M - 1}
    \Pr\left({\boldsymbol G}_u \to {\boldsymbol G}_t\right)
\end{equation}
where ${\rm Pr} \left({\boldsymbol G}_u \to {\boldsymbol
G}_t\right)$ is the pairwise error probability (PEP) of the OSTBC,
i.e. the probability of detecting ${\boldsymbol G}_t$ when
${\boldsymbol G}_u$ is transmitted.

The closed-form solution for the PEP of ML decoding for OSTBCs
with arbitrary constituent signals in the Rayleigh fading channel
is well known (e.g., see \cite{JafarkhaniH05} and the references
therein). Indeed, the PEP is calculated as the expectation of
$\Pr\left({\boldsymbol G}_u \to {\boldsymbol G}_t\, \vert \,
{\boldsymbol H}\right)$ over ${\boldsymbol H}$, where
${\boldsymbol H} \triangleq [{\boldsymbol h}_1 \comdots
{\boldsymbol h}_{N_R}]$ is the matrix of channel coefficients.
Using our notation for the equivalent Euclidean codes, the PEP of
ML decoding for OSTBCs with arbitrary constituent signals can be
obtained as
\begin{subequations}\label{eq15}
    \begin{equation}
        \Pr\left({\boldsymbol G}_u \to {\boldsymbol G}_t\right)
        = \frac{1}{2} - \frac{\mu_{u, t}}{2} \sum_{r = 0}^{K - 1}
        \binom{2 r}{\!r} \left(\frac{1 - \mu_{u, t}^2}{4}\right)^{\!r}
    \end{equation}
    where
    \begin{equation}\label{kdef}
        K \triangleq N_T N_R
    \end{equation}
    and where
    \begin{equation}
        \mu_{u, t} \triangleq \sqrt{\frac{\tilde{d}_{u, t,
        {\rm EC}}^{\,2} \, \bar{\gamma}_c}{4 + \tilde{d}_{u, t,
        {\rm EC}}^{\,2} \, \bar{\gamma}_c}}
    \end{equation}
    where
    \begin{equation}
        \bar{\gamma}_c \triangleq \frac{\rho \bar{E}_{\rm EC}}{N_0}
    \end{equation}
\end{subequations}
is the average received Euclidean code-to-noise ratio (cf.
\eqref{avgsnr}).

To analyze the asymptotic behavior of the PEP \eqref{eq15}, the
following new lemma is useful.

{\bf Lemma 1:} {\it The PEP of the OSTBC \eqref{eq15} satisfies
the following identity}
\begin{equation}\label{eq17}
    \frac{1}{2} - \frac{\mu_{u, t}}{2} \sum_{r = 0}^{K - 1}
    \binom{2 r}{\!r} \left(\frac{1 - \mu_{u, t}^2}{4}\right)^{\!r}
    = \left(\frac{1 - \mu_{u, t}}{2}\right)^{\!K} \,
    \sum_{r = 0}^{K - 1} \binom{K - 1 + r}{r} \left(\frac{1 +
    \mu_{u, t}}{2}\right)^{\!r}.
\end{equation}

{\bf Proof:} Substituting $z = (1 + \mu_{u, t}) / 2$ into the
combinatorial identity \cite[eq. (5.138)]{GrahamRL94}
\begin{equation}\label{eq16}
    1 + \frac{1 - 2 z}{2 - 2 z} \sum_{r = 1}^n \binom{2 r}{r}
    \left(z (1 - z)\right)^r = (1 - z)^n \sum_{r = 0}^n
    \binom{n + r}{r} z^r
\end{equation}
we immediately obtain \eqref{eq17}. $\hfill\square$

Substituting \eqref{eq17} in \eqref{eq14a} yields the union bound
for the SER of OSTBCs with uniform signal constellations in the
Rayleigh fading channel in the form
\begin{equation}\label{eq18}
    {\Pr}_{s, \rm{OSTBC}} \leq  \sum_{t = 0, t \neq u}^{M - 1}
    \left(\frac{1 - \mu_{u, t}}{2}\right)^{\!K} \,
    \sum_{r = 0}^{K - 1} \binom{K - 1 + r}{r} \left(\frac{1 +
    \mu_{u, t}}{2}\right)^{\!r}
\end{equation}
where $K$ is given by \eqref{kdef}.

\subsection{Optimality of the OSTBC Signal Constellation:
Large SNR}
In the case of large SNR, i.e. $\bar{\gamma}_c \gg 1$,
approximating the terms $(1 - \mu_{u, t}) / 2$ and $(1 + \mu_{u,
t}) / 2$ using the first component of their Taylor series yields
$(1 - \mu_{u, t}) / 2 \approx 1 / (\tilde{d}_{u, t, {\rm
EC}}^{\,2} \, \bar{\gamma}_c)$ and $(1 + \mu_{u, t}) / 2 \approx
1$. Furthermore, using the following combinatorial expression
\cite[eq.~(14.4-17)]{ProakisJG08}
\begin{equation}
    \sum_{r = 0}^{K - 1} \binom{K - 1 + r}{r} = \binom{2 K - 1}{K}
\end{equation}
we can approximate \eqref{eq18} for the case of large SNR as
\begin{subequations}\label{eq1920}
    \begin{equation}\label{eq19}
        P_{s, \rm{OSTBC}} \leq C_{\rm EC}(K) \binom{2 K
        - 1}{K} \bar{\gamma_c}^{-K}
    \end{equation}
    where $K$ is defined by \eqref{kdef} and where
    \begin{equation}\label{eq20}
    C_{\rm EC}(K) \triangleq \sum_{t = 0, t \neq u}^{M
    - 1} \tilde{d}_{u, t, {\rm EC}}^{\,-2 K}
    \end{equation}
\end{subequations}
which is called here the normalized distance spectrum constant
(NDSC) of the OSTBC. It is interesting to note that the NDSC is a
fixed parameter of an OSTBC for a given $K$. The NDSC is defined
only by the distance properties of the Euclidean code equivalent
to the OSTBC and it does not depend on SNR. Therefore, we can say
that a Euclidean code with minimal $C_{\rm EC}(K)$ among all
Euclidean codes with the same $M$, $K$, and dimensionality $n$ is
optimal in the sense that it provides the smallest SER at large
SNR. Here, identical dimensionality ensures the same requirement
for time/frequency resources and the same required number of
transmitted bits per dimension, needed for fair comparison of the
SER. The following theorem gives a more precise statement of the
optimality.

{\bf Theorem 2:} {\it For a quasistatic fading channel, large SNR
$\bar{\gamma}_c \gg 1$, and a given $K$, an OSTBC with cardinality
$M$ is optimal if and only if the Euclidean code equivalent to
this OSTBC has the minimal NDSC \eqref{eq20} among all Euclidean
codes with the same $M$ and dimensionality.}

{\bf Proof:} Both necessity and sufficiency follow directly from
\eqref{eq1920}. If an OSTBC is optimal, it has the minimal NDSC.
Otherwise, a code with a smaller NDSC achieves smaller SER
according to \eqref{eq1920}. Conversely, if an OSTBC has the
minimal NDSC, it is not outperformed by any other OSTBC, as the
latter has an equal or larger NDSC, and thus SER, based on
\eqref{eq1920}.$\hfill\square$

The following, perhaps obvious, but important corollary follows
from Theorem 2.

{\bf Corollary 1:} For a quasistatic fading channel, large SNR,
and given $K$, an OSTBC signal constellation is optimal if and
only if the Euclidean code equivalent to this OSTBC has the
minimal NDSC \eqref{eq20} among all Euclidean codes with the same
$M$ and dimensionality.

This corollary formulates the general criterion for designing
optimal OSTBC signal constellations on quasistatic fading
channels. To the best of the authors' knowledge, this is a new
general design criterion for optimal OSTBC signal constellations.
Moreover, as also follows from \eqref{eq20} and Theorem 2, the
optimality of the Euclidean code equivalent to an OSTBC for a
given number of receiving antennas, $N_R$, is not a sufficient
condition for the optimality of the same code for a different
number of receiving antennas. This is due to the nonlinear
behavior of the NDSC \eqref{eq20} with respect to $N_R$. The
following remark formalizes the novelty of the results given
above.

{\bf Remark 2:} Methods of design for Euclidean codes with minimal
NDSC are not known. Also, the results embodied in \eqref{eq1920}
have not appeared before in the literature in the context of
OSTBCs. As a result, Euclidean codes satisfying the conditions of
Theorem 2 have not yet been investigated for any OSTBC. Moreover,
there is no regular method of design for any class of Euclidean
codes which are optimal according to any design criterion.

\subsection{Optimality of the OSTBC Signal Constellation:
Large SNR $\bar{\gamma }_c \gg 1$ and for a Large Number of
Antennas $N_T, N_R \gg 1$} In this case, the NDSC \eqref{eq20} can
be approximated as
\begin{equation}\label{ndscapprox}
    C_{\rm EC}(K) \approx N_{\tilde{d}_{\min, {\rm EC}}} \,
    \tilde{d}_{\min, {\rm EC}}^{\,-2 K}
\end{equation}
where $\tilde{d}_{\min, {\rm EC}}$ is the minimal normalized
Euclidean distance of the Euclidean code equivalent to the OSTBC
and where $N_{\tilde{d}_{\min, {\rm EC}}}$ is the number of
codewords with the minimal distance $\tilde{d}_{\min, {\rm EC}}$.
Approximation \eqref{ndscapprox} simplifies bound \eqref{eq19} to
\begin{equation}\label{eq21}
    P_{s, \rm{OSTBC}} \leq N_{\tilde{d}_{\min, {\rm EC}}} \,
    \tilde{d}_{\min, {\rm EC}}^{\,-2 K} \binom{2 K - 1}{K}
    \bar{\gamma}_c^{-K}.
\end{equation}

The bound \eqref{eq21} is well known (e.g., see
\cite{BrehlerM01_TInfo}), and thus, can serve as a check on our
previous derivations. However, we use this bound here to derive
existence conditions for OSTBCs based on their connection with the
equivalent Euclidean codes. To the best of the authors' knowledge,
such discussion has not appeared in the literature before. Note
from \eqref{eq21} that the dominant parameter for OSTBC optimality
in the case of large SNR and for a large number of antennas is the
minimal distance of the equivalent Euclidean code.

Although the aforementioned design criterion based on \eqref{eq21}
is known, what has not been exploited before is that such a
criterion coincides with the standard one for the error correcting
codes optimal for the Gaussian channel. Thus, results for the
optimal Euclidean codes known from the classic theory of error
correcting coding can be used to define the existence conditions
of the optimal OSTBC for large SNR and for a large number of
antennas.

An interesting special case of Euclidean codes is \textit{a
spherical code}, for which every symbol of the code has the same
norm \cite{EricsonT01}. Since we are interested in designing
optimal OSTBCs, the notion of optimality for spherical codes is of
importance. The \textit{optimal spherical code} \cite{EricsonT01}
is the code with the maximal minimum normalized Euclidean distance
$\tilde{d}_{\min, {\rm EC}}$ among all spherical codes with the
same cardinality $M$ and dimensionality $n$. Note that the
Euclidean codes equivalent to OSTBCs with constant energy
constituent signals belong to the class of spherical codes. This
leads to interesting connections between OSTBCs and error
correcting codes. Particularly, the bounds obtained for the
spherical codes can be used to define parameters of the
asymptotically optimal OSTBCs. Some of the strongest and deepest
results on the existence conditions of spherical codes with small
dimensionality and squared minimal Euclidean distance $0 <
\tilde{d}_{\min}^2 \le 4$ were obtained by Rankin
\cite{RankinRA55_GMA} (see also \cite[Ch.~1.4]{EricsonT01}) and
Coxeter-B\"{o}r\"{o}czky \cite[p.~28]{ConwayJ98}. Based on the
results of Rankin and Coxeter-B\"{o}r\"{o}czky for spherical
codes, the following bounds for asymptotically optimal signal
constellations of OSTBCs can be formulated for the case of large
SNR and for a large number of antennas.

{\bf Theorem 3 {\it(Similar to the Coxeter-B\"{o}r\"{o}czky
bound)}:} {\it For quasistatic fading channels at large SNR and
for a large number of antennas, any asymptotically optimal OSTBC
that uses constituent signals with equal energies in the $n \ge 2$
dimensional Euclidean space, satisfies the conditions}
\begin{equation}
    \tilde{d}_{\min} = 2 \sin\frac{\alpha}{2}
\end{equation}
{\it and}
\begin{equation}
    M \le \frac{2F_{n - 1}(\alpha)}{F_n(\alpha)}
\end{equation}
{\it where $F_n(\alpha)$ is the Schl\"{a}fli's function defined as}
\begin{equation}
    F_n(\alpha)=\frac{2}{\pi} \int_{\frac{1}{2}
    \mathrm{arcsec}(n - 1)}^\alpha F_{n - 2}(\beta) \mathrm{d}\alpha
\end{equation}
{\it where $\sec(\mbox{2}\beta) = \sec(\mbox{2}\alpha) - 2$,
$F_0(\alpha) = F_1(\alpha) = 1$, $0 < \alpha \le \pi$.}

{\bf Proof:} An OSTBC with equal-energy constituent signals
corresponds to a Euclidean spherical code. Under the asymptotic
hypotheses of the theorem, the optimality of the OSTBC corresponds
to the maximality of the minimum Euclidean distance of the
equivalent spherical code. This maximality condition is satisfied
under the claims of the theorem, based on \cite[p.~28]{ConwayJ98}.

{\bf Theorem 4 {\it(Similar to Rankin's first bound)}:} {\it For
quasistatic fading channels at large SNR and for a large number of
antennas, any asymptotically optimal OSTBC that uses constituent
signals with equal energies satisfies the inequality}
\begin{equation}\label{eq22}
    \tilde{d}_{\min}^2 \le \frac{2 M}{M - 1}.
\end{equation}

{\bf Proof:} See the proof of Theorem 3, and we also refer to
\cite{RankinRA55_GMA} and \cite[Ch.~1.4]{EricsonT01}.

{\bf Remark 3:} An interesting fact about the bound \eqref{eq22}
is that it does not depend on the dimensionality of the code.

{\bf Theorem 5 {\it(Similar to Rankin's second bound)}:} {\it For
quasistatic fading channels at large SNR and for a large number of
antennas, the largest $M$ of an OSTBC that uses constituent
signals with equal energies satisfies the inequality}
\begin{equation}
    M \le n + 1
\end{equation}
{\it for $2 < \tilde{d}_{\min}^2 \le 4$.}

{\bf Proof:} See the proof of Theorem 3, and we also refer to
\cite{RankinRA55_GMA} and \cite[Ch.~1.4]{EricsonT01}.

{\bf Theorem 6 {\it(Similar to Rankin's third bound)}:} {\it For
quasistatic fading channels at large SNR and for a large number of
antennas, the largest $M$ of an OSTBC that uses constituent
signals with equal energies satisfies the inequality}
\begin{equation}\label{eq24}
    M \le 2 n
\end{equation}
{\it for $\tilde{d}_{\min}^2 = 2$.}

{\bf Proof:} See the proof of Theorem 3, and we also refer to
\cite{RankinRA55_GMA} and \cite[Ch.~1.4]{EricsonT01}.

The importance of Theorems 3--6 is especially stressed by the fact
that these theorems provide the only known general bounds on $M$
as existence conditions for asymptotically optimal OSTBCs using
constituent signals with equal energies, assuming coherent
receivers, quasistatic fading channels, large SNR, and a large
number of antennas.

{\bf Remark 4:} Although OSTBCs are connected now to spherical
codes, it is still worth noting that regular methods for designing
spherical codes with constituent modulated signals are not known.
Thus, the code design problem is still not a simple problem, but
such connections allow us to exploit some results on the design of
spherical codes, such as a number of results summarized in
\cite{EricsonT01}. Moreover, an approach based on the theory of
group codes \cite{ProakisJG08, ZetterbergLH77_TCom} can also be
useful, although methods for regular design of group codes with
optimal distance properties are not known either. A possible
undesirable consequence of considering group codes is that the
constituent signals of these codes have symmetric properties; this
is a severe restriction for code design and can result in
nonoptimal codes. Finally, it is noteworthy that some useful
properties of group codes suitable for the signal constellations
of OSTBCs have been exploited in the OSTBC literature (e.g., see
the research works on unitary code design \cite{HughesBL03_TInfo,
LiangXB02_TInfo}).

\subsection{New Asymptotically Optimal $M = 8$ and $M = 16$
Biorthogonal Signal Constellations for the Alamouti
OSTBC}\label{codedesign} As an example of code design based on our
studies in this section, we consider biorthogonal spherical codes.
Indeed, biorthogonal spherical codes can be constructed for almost
any multidimensional space \cite{EricsonT01}.

\begin{figure}
    \psfrag{S0, S1}{$\boldsymbol{s}_0, \boldsymbol{s}_1$}
    \psfrag{S2, S3}{$\boldsymbol{s}_2, \boldsymbol{s}_3$}
    \psfrag{S6, S7}{$\boldsymbol{s}_6, \boldsymbol{s}_7$}
    \psfrag{S4, S5}{$\boldsymbol{s}_4, \boldsymbol{s}_5$}
    \psfrag{S0, S6}{$\boldsymbol{s}_0, \boldsymbol{s}_6$}
    \psfrag{S2, S4}{$\boldsymbol{s}_2, \boldsymbol{s}_4$}
    \psfrag{S1, S7}{$\boldsymbol{s}_1, \boldsymbol{s}_7$}
    \psfrag{S3, S5}{$\boldsymbol{s}_3, \boldsymbol{s}_5$}
    \psfrag{S1u}{$s_{1, u}$}
    \psfrag{S2u}{$s_{2, u}$}
    \centering
    \includegraphics{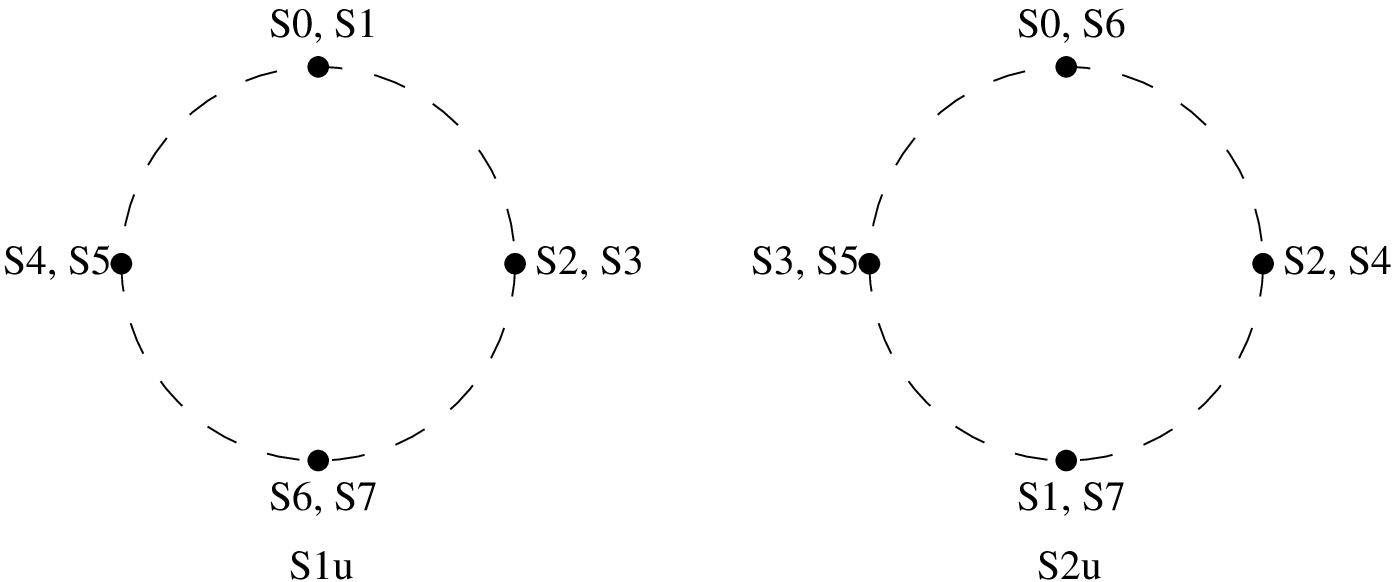}
    \vspace{-5pt}
    \caption{Optimal signal constellation of a 4-D biorthogonal
    code with $M = 8$ for Alamouti's code.}
    \label{fig_SigAlam4dim}
\end{figure}

\begin{figure}
    \psfrag{S0, S1, S2, S3}[][]{$\boldsymbol{s}_0, \boldsymbol{s}_1, \boldsymbol{s}_2, \boldsymbol{s}_3$}
    \psfrag{S12, S13, S14, S15}[][]{$\boldsymbol{s}_{12}, \boldsymbol{s}_{13}, \boldsymbol{s}_{14}, \boldsymbol{s}_{15}$}
    \psfrag{S8, S9, S10, S11}[l][l]{\parbox{40pt}{\centering$\boldsymbol{s}_8$, $\boldsymbol{s}_9$, $\boldsymbol{s}_{10}$, $\boldsymbol{s}_{11}$}}
    \psfrag{S4, S5, S6, S7}[l][l]{\parbox{40pt}{\centering$\boldsymbol{s}_4$, $\boldsymbol{s}_5$, $\boldsymbol{s}_6$, $\boldsymbol{s}_7$}}
    \psfrag{S0, S1, S12, S13}[][]{$\boldsymbol{s}_0, \boldsymbol{s}_1, \boldsymbol{s}_{12}, \boldsymbol{s}_{13}$}
    \psfrag{S2, S3, S14, S15}[][]{$\boldsymbol{s}_2, \boldsymbol{s}_3, \boldsymbol{s}_{14}, \boldsymbol{s}_{15}$}
    \psfrag{S5, S6, S8, S11}[l][l]{\parbox{40pt}{\centering$\boldsymbol{s}_5$, $\boldsymbol{s}_6$,\\$\boldsymbol{s}_8$, $\boldsymbol{s}_{11}$}}
    \psfrag{S4, S7, S9, S10}[l][]{\parbox{40pt}{\centering$\boldsymbol{s}_4$, $\boldsymbol{s}_7$,\\$\boldsymbol{s}_9$, $\boldsymbol{s}_{10}$}}
    \psfrag{S0, S2, S12, S14}[][]{$\boldsymbol{s}_0, \boldsymbol{s}_2, \boldsymbol{s}_{12}, \boldsymbol{s}_{14}$}
    \psfrag{S1, S3, S13, S15}[][]{$\boldsymbol{s}_1, \boldsymbol{s}_3, \boldsymbol{s}_{13}, \boldsymbol{s}_{15}$}
    \psfrag{S5, S7, S9, S11}[l][l]{\parbox{40pt}{\centering$\boldsymbol{s}_5$, $\boldsymbol{s}_7$, $\boldsymbol{s}_9$, $\boldsymbol{s}_{11}$}}
    \psfrag{S4, S6, S8, S10}[l][l]{\parbox{40pt}{\centering$\boldsymbol{s}_4$, $\boldsymbol{s}_6$, $\boldsymbol{s}_8$, $\boldsymbol{s}_{10}$}}
    \psfrag{S0, S3, S13, S14}[][]{$\boldsymbol{s}_0, \boldsymbol{s}_3, \boldsymbol{s}_{13}, \boldsymbol{s}_{14}$}
    \psfrag{S1, S2, S12, S15}[][]{$\boldsymbol{s}_1, \boldsymbol{s}_2, \boldsymbol{s}_{12}, \boldsymbol{s}_{15}$}
    \psfrag{S6, S7, S10, S11}[l][l]{\parbox{40pt}{\centering$\boldsymbol{s}_6$, $\boldsymbol{s}_7$,\\$\boldsymbol{s}_{10}$, $\boldsymbol{s}_{11}$}}
    \psfrag{S4, S5, S8, S9}[l][]{\parbox{40pt}{\centering$\boldsymbol{s}_4$, $\boldsymbol{s}_5$,\\$\boldsymbol{s}_8$, $\boldsymbol{s}_9$}}
    \psfrag{S1u}{$s_{1, u}$}
    \psfrag{S2u}{$s_{2, u}$}
    \centering
    \includegraphics{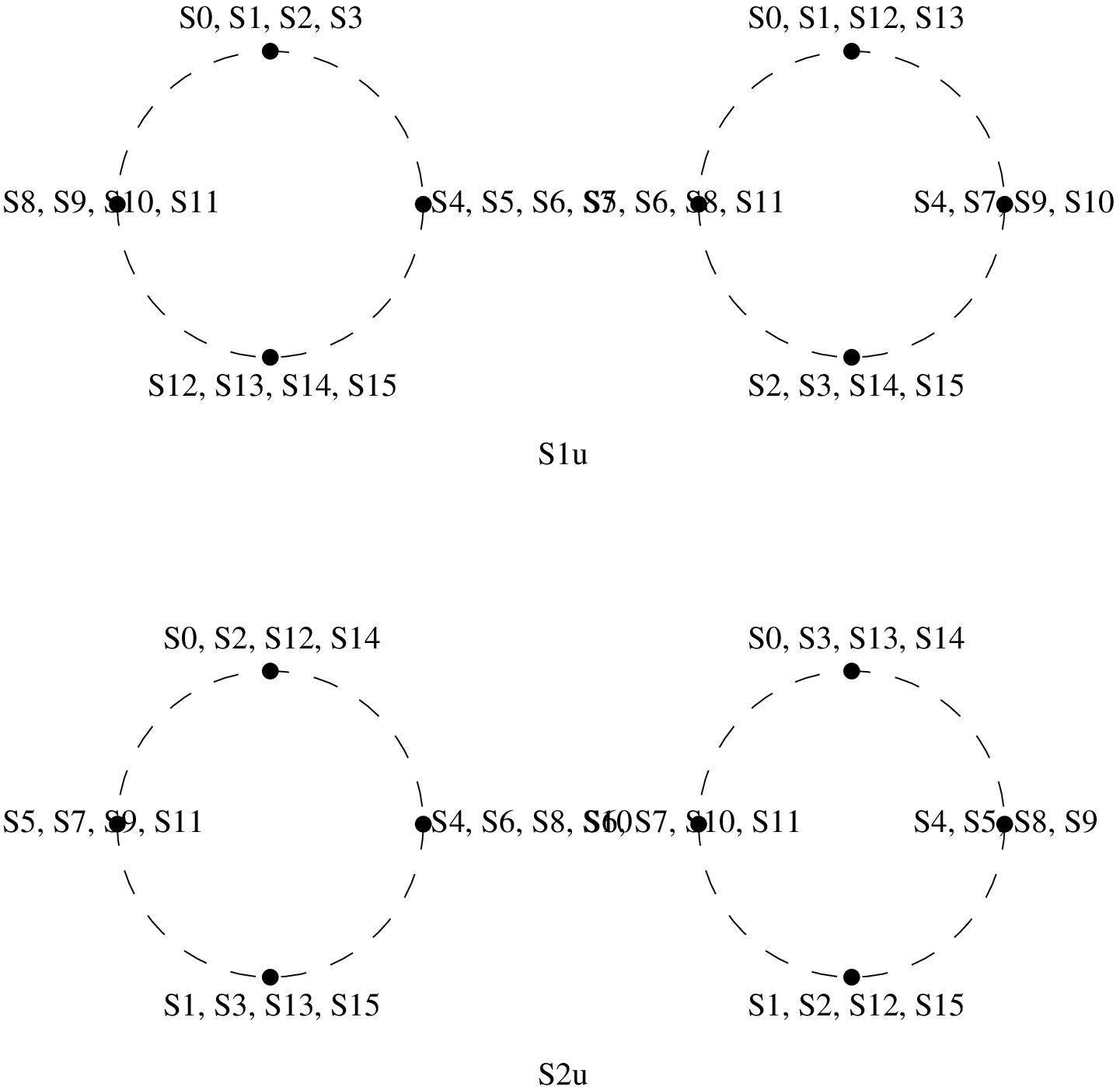}
    \caption{Optimal signal constellation of an 8-D biorthogonal
    code with $M = 16$ for Alamouti's code.}
    \label{fig_SigAlam8dim}
\end{figure}

\begin{figure}
    \psfrag{S0}{\large$\boldsymbol{s}_0$}
    \psfrag{S1}{\large$\boldsymbol{s}_1$}
    \psfrag{S2}{\large$\boldsymbol{s}_2$}
    \psfrag{S3}{\large$\boldsymbol{s}_3$}
    \psfrag{S4}{\large$\boldsymbol{s}_4$}
    \psfrag{S5}{\large$\boldsymbol{s}_5$}
    \psfrag{S6}{\large$\boldsymbol{s}_6$}
    \psfrag{S7}{\large$\boldsymbol{s}_7$}
    \psfrag{sqrt}{\large$\sqrt{2}$}
    \centering
    \includegraphics[scale=.9]{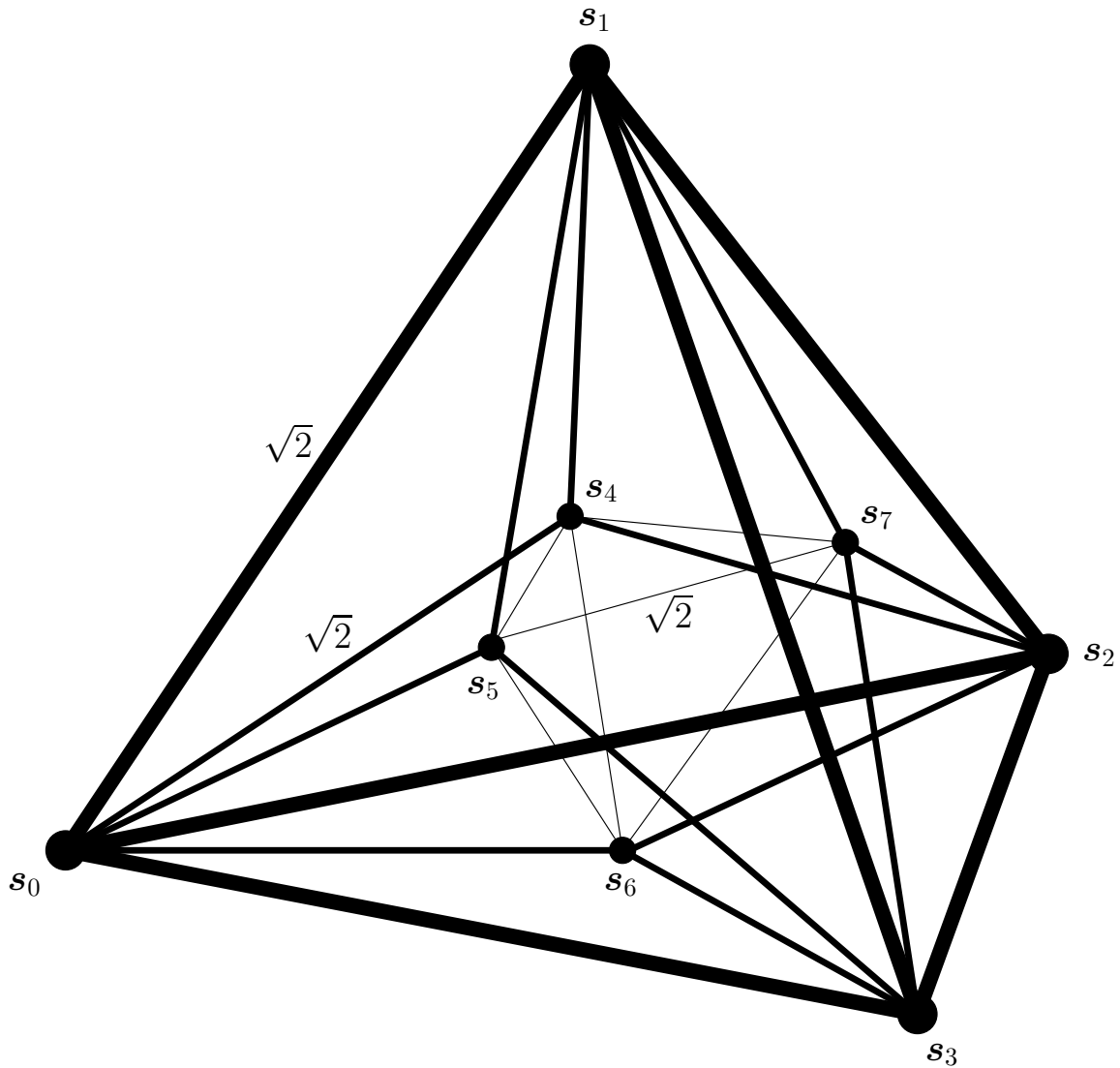}
    \caption{The Schlegel diagram of the hexadecachoron (16-cell)
    in the 4-D space [after http://en.wikipedia.org/wiki/16-cell]
    showing a geometrical representation of the optimal signal
    constellation with $M = 8$ (the biorthogonal spherical code
    given in Fig. \ref{fig_SigAlam4dim}) for Alamouti's code.}
    \label{fig_VecAlam4dim}
\end{figure}

Consider a 4-D biorthogonal code with $M = 8$ and an 8-D
biorthogonal code with $M = 16$. Such codes satisfy the upper
bound \eqref{eq24} with equality. Therefore, the signal
constellations of the new biorthogonal spherical codes depicted in
Figs. \ref{fig_SigAlam4dim} and \ref{fig_SigAlam8dim} based on
QPSK signaling can serve as examples of new asymptotically optimal
signal constellations for Alamouti's codes with $M = 8$ and $M =
16$. The spectrum and graphical representation of the code with $M
= 8$ are given in Table \ref{tab_alambio} and Fig.
\ref{fig_VecAlam4dim}, respectively. The spectrum of the code with
$M = 16$ is similar to that of the code with $M = 8$; only the
number $6$ in Table \ref{tab_alambio} changes to $14$. The signal
coordinate diagram of the code with $M = 16$ (in 8-D space) has
not been depicted as it is cumbersome. Note from Figs.
\ref{fig_SigAlam4dim} and \ref{fig_SigAlam8dim} that codewords
$\boldsymbol{s}_0 \comdots \boldsymbol{s}_{M / 2 - 1}$ are
orthogonal, and are respectively the complements of
$\boldsymbol{s}_{M - 1} \comdots \boldsymbol{s}_{M / 2}$ to ensure
Grey mapping. Performance simulations for these two codes are
presented in Section \ref{numresults}.

\begin{table}
\caption{Normalized Distance Spectrum of the Biorthogonal
Spherical Code Equivalent to Alamouti's Code\protect\\With
Biorthogonal 4-D Constituent Signal Constellation and $M = 8$}
\vspace{-20pt}
\begin{center}
    \begin{tabular}{lcc}
        \toprule\addlinespace[1pt]
        \rule{0pt}{13pt} Normalized Euclidean distance
        & ~~$\sqrt{2}$~~ & ~~$2$~~ \\\addlinespace[1pt]
        \hline
        \rule{0pt}{13pt} Number of codewords & ~~$6$~~ & ~~$1$~~ \\
        \bottomrule
    \end{tabular}
\end{center}
\label{tab_alambio}
\end{table}

\section{OSTBC Performance Analysis}
The existing results on OSTBC performance analysis (see
\cite{BauchG01_AnnTelecom, GaoC02_VTC, Gharavi-AlkhansariM04_TCom,
ZhangH05_TWireless, LuH03_TInfo, BehnamfarF05_TCom,
SandhuS01_ICASSP, BrehlerM01_TInfo} and the references therein)
aim at deriving exact expressions only for the SER of the
constituent signals of the OSTBCs, while there are no results on
the SER of an OSTBC in the sense of the probability that a
codeword (code matrix) is transmitted but another codeword is
detected. However, it is the latter SER for all types of
modulation and coding, including orthogonal space-time coding,
that is a common and important performance evaluation measure in
communication systems.

Performance analysis of orthogonal space-time coding MIMO
communication systems in a fading channel can be performed using
the proposed equivalent model for MIMO systems given in Fig.
\ref{fig_EquiModel}. This model is connected to the classic
receiver diversity system (e.g., see \cite{ProakisJG08} and
\cite[Fig. 1]{SimonMK98_Proc}), which has long been of interest.
However, the significant difference between our model in Fig.
\ref{fig_EquiModel} and the classic receiver diversity system is
that our model is, in fact, a form of receiver diversity of the
block coded signals. This difference is especially useful from the
performance analysis point of view.

\subsection{Methodology for the General Case of Arbitrary
Constituent Signals} Using the model in Fig. \ref{fig_EquiModel},
the bit error rate (BER), i.e. $\Pr_b$, and SER, i.e. $\Pr_s$, of
an OSTBC with arbitrary constituent signals over the Rayleigh
fading channel can be evaluated based on the classic approach of
estimating the error performance of a digital communication system
over fading channels. That is, the BER and SER of the equivalent
Euclidean code are evaluated by statistically averaging the
conditional BER $\Pr_b^{\rm EC}(\gamma_b)$ and SER $\Pr_s^{\rm
EC}(\gamma_b)$ at the output of a coherent receiver of this code
for the Gaussian channel over the joint probability density
function (PDF) of the fading amplitudes $f_{\gamma_b}(\gamma_b)$
as
\begin{equation}\label{eq25}
    {\Pr}_{b, \rm{OSTBC}} = \int_0^\infty {\Pr}_b^{\rm EC}
    (\gamma_b) \, f_{\gamma_b}(\gamma_b) \, \mathrm{d}\gamma_b
\end{equation}
\begin{equation}\label{eq26}
    {\Pr}_{s, \rm{OSTBC}} = \int_0^\infty {\Pr}_s^{\rm EC}
    (\gamma_b) \, f_{\gamma_b}(\gamma_b) \, \mathrm{d}\gamma_b
\end{equation}
where $\gamma_b$ is the total instantaneous SNR per bit at the
output of the ML receiver given by
\begin{equation}\label{eq27}
    \gamma_b \triangleq \sum_{j = 1}^{N_R} {\gamma_j}
\end{equation}
where $\gamma_j $ is the instantaneous SNR per bit in $j$th
channel.

Towards evaluating \eqref{eq25} and \eqref{eq26}, it is first
required to obtain the PDF of the combined fading coefficient
$f_{\gamma_b}(\gamma_b)$, conditional BER ${\Pr}_b^{\rm
EC}(\gamma_b)$, and SER ${\Pr}_s^{\rm EC}(\gamma_b)$ of the
equivalent Euclidean code on the Gaussian channel. As also follows
from \eqref{eq25} and \eqref{eq26}, the main problem of
performance analysis of OSTBCs based on the proposed model in Fig.
\ref{fig_EquiModel} can be reduced to the evaluation of BER/SER of
the corresponding Euclidean code over the channel with Gaussian
noise. For example, the problem is reduced to the BER/SER
evaluation of a 4-D Euclidean code in the case of the canonical
Alamouti code with 2-D constituent signals and to the evaluation
of a 6-D Euclidean code in the case of the rate 3/4 OSTBC with 2-D
constituent signals. Although this methodology for exact BER/SER
evaluation for the multidimensional signal constellations of
interest is straightforward after the model in Fig.
\ref{fig_EquiModel} is introduced, it cannot be found in the
available literature and appears for the first time here. Another
general methodology for performance analysis of OSTBCs with
arbitrary constituent signals has been formulated in
\cite{ZhangH05_TWireless}. However, our methodology based on the
classic performance analysis approach and the equivalent model in
Fig. \ref{fig_EquiModel} is more straightforward and appears to be
significantly simpler than the approach of
\cite{ZhangH05_TWireless}.

\subsection{BER and SER of Alamouti's Code With Constituent
BPSK Signals Over the Rayleigh Fading
Channel}\label{serberanalysis} As an example of applying our
performance analysis methodology, we  derive a closed-form
solution for the BER and SER of the Alamouti OSTBC with
constituent BPSK signals. The signal coordinate diagram of the
equivalent Euclidean code for Alamouti's code with constituent
BPSK signals is given in Fig. \ref{fig_VecAlamBPSK}. Note that
this diagram coincides with the signal coordinate diagram of QPSK
with Grey mapping. Therefore, the Alamouti scheme with constituent
BPSK signals corresponds to the receiver diversity scheme of Fig.
\ref{fig_EquiModel} with QPSK signaling and with the ML receiver
simplifying to the maximal-ratio combining receiver.

The expressions for $\Pr_b^{\rm EC}(\gamma_b)$ and $\Pr_s^{\rm EC}
(\gamma_b)$ of a coherent receiver for QPSK with Grey mapping in
the Gaussian channel are well known and can be found, for example,
in \cite{JafarkhaniH05} as
\begin{equation}\label{eq28}
    {\Pr}_b^{\rm EC}(\gamma_b) = Q \!\left(\!\sqrt{2
    \gamma_b}\right)
\end{equation}
\begin{equation}\label{eq29}
    {\Pr}_s^{\rm EC}(\gamma_b) = 2 \, Q\!\left(\!\sqrt{2
    \gamma_b}\right) - Q^2\!\left(\!\sqrt{2\gamma_b}\right)
\end{equation}
where $Q(x) = \frac{1}{\sqrt{2 \pi}} \int_x^\infty
\mathrm{e}^{-t^2 / 2} \, \mathrm{d}t$ is the Gaussian
$Q$-function. Also, the average SNR per bit is
\begin{equation}
    \gamma_b = \frac{\bar{E} \, \eta}{N_T N_0 \log_2 M}
\end{equation}
where, as follows from the model in Fig. \ref{fig_EquiModel},
$\eta \triangleq \sum_{j = 1}^{N_R} \left\|c_j\right\|^2 = \sum_{i
= 1}^{N_T} \sum_{j = 1}^{N_R} \|h_{i, j}\|^2$, $\bar{E} \triangleq
2 E_b$, and $M = 4$ since the signaling is quaternary. Then, the
average SNR per bit can be expressed as
\begin{equation}\label{eq30}
    \gamma_b = \frac{E_b}{N_T N_0} \sum_{i = 1}^{N_T}
    \sum_{j = 1}^{N_R} \|h_{i,j}\|^2.
\end{equation}

It has been shown in \cite{ProakisJG08} (see also
\cite{BauchG01_AnnTelecom}) that the PDF of the average per bit
SNR \eqref{eq30} is given as
\begin{equation}\label{eq31}
    f_{\gamma_b}(\gamma) = \frac{1}{(K - 1)! \bar{\gamma_b}^K}
    \, \gamma^{K - 1} \, \mathrm{e}^{-\gamma / \bar{\gamma_b}}
\end{equation}
where $K$ is defined by \eqref{kdef} and where $\bar{\gamma}_b
\triangleq E_b / N_T N_0$. Substituting \eqref{eq28} and
\eqref{eq31} into \eqref{eq25}, the average BER of the Alamouti
code with constituent BPSK signals can be expressed as
\begin{equation} \label{eq32}
    {\rm Pr}_b = \int_0^\infty \frac{1}{(K - 1)!
    \bar{\gamma_b}^K} \, \gamma^{K - 1} \, \mathrm{e}^{-\gamma
    / \bar{\gamma_b}} \, Q\!\left(\!\sqrt{2 \gamma}\right) \,
    \mathrm{d}\gamma.
\end{equation}
Moreover, after some computations, it can be derived that
\begin{equation}\label{eq33}
    {\Pr}_b = \frac{1}{2} - \frac{\mu_b}{2} \sum_{r = 0}^{K - 1}
    \binom{2 r}{r} \left(\frac{1 - \mu_b^2}{4}\right)^r
\end{equation}
where $\mu_b \triangleq \sqrt{\bar{\gamma}_b / (1 +
\bar{\gamma}_b)}$. Note that the solution \eqref{eq33} is not new
and has been derived by Bauch et al. in \cite{BauchG01_AnnTelecom}
based on the SISO model and later also verified in
\cite{Gharavi-AlkhansariM04_TCom, ZhangH05_TWireless,
LuH03_TInfo}.

The average SER of Alamouti's code with BPSK constituent signals
has not been obtained previously. Substituting \eqref{eq29} and
\eqref{eq31} into \eqref{eq26}, the average SER can be expressed
as
\begin{equation}\label{eq34}
    {\Pr}_s = \int_0^\infty \frac{1}{\left(K - 1\right)! \,
    \bar{\gamma_b}^K} \, \gamma^{K - 1} \, \mathrm{e}^{-\gamma
    / \bar{\gamma}_b} \left(2 Q\!\left(\!\sqrt{2\gamma}\right) -
    Q^2\!\left(\!\sqrt{2\gamma}\right)\right) \, \mathrm{d}\gamma.
\end{equation}
Moreover, using \cite[eqs. (2) and (6)]{SimonMK98_Proc} and
performing some computations, the expression \eqref{eq34} can be
rewritten as
\begin{equation}\label{eq35}
    {\rm Pr}_s = \frac{2}{\pi} \int_0^{\pi / 4} \left(
    \frac{\cos^2 \theta}{\cos^2 \theta + \bar{\gamma}_b}
    \right)^{\!K} \! \mathrm{d}\theta + \frac{1}{\pi}
    \int_0^{\pi / 4} \left(\frac{\sin^2 \theta}{\sin^2
    \theta + \bar{\gamma}_b}\right)^{\!K} \! \mathrm{d}
    \theta.
\end{equation}
To the best of the authors' knowledge, \eqref{eq35} is a new
expression for the SER of Alamouti's code with the constituent
BPSK signals. Moreover, this is the only available exact
expression for the SER of any OSTBC. All other known results are
for the SER of the constituent signals of the OSTBC, that is,
obviously, not the same and less descriptive of system
performance.

\section{Numerical Examples}\label{numresults}
New asymptotically optimal signal constellations for the Alamouti
OSTBC were found in Section \ref{codedesign}. The performances of
these codes represent best cases of interest and are determined by
simulation in this section.

\begin{figure}
    \psfrag{SNR, dB}[][]{\raisebox{-20pt}{\small Average SNR, dB}}
    \psfrag{BER}[][]{\raisebox{25pt}{\small BER}}
    \psfrag{Alamouti QPSK, analysis}{\small Alamouti QPSK, analysis}
    \psfrag{Alamouti 4-D, simulation}{\small Alamouti 4-D, simulation}
    \psfrag{Rate 3/4 OSTBC, analysis}{\small Rate 3/4 OSTBC, analysis}
    \psfrag{NR = 1}{\small$N_R = 1$}
    \psfrag{NR = 2}{\small$N_R = 2$}
    \psfrag{NR = 4}{\small$N_R = 4$}
    \psfrag{0}[][]{\small$0$}
    \psfrag{2}[][]{\small$2$}
    \psfrag{4}[][]{\small$4$}
    \psfrag{6}[][]{\small$6$}
    \psfrag{8}[][]{\small$8$}
    \psfrag{10}[][]{\small$10$}
    \psfrag{12}[][]{\small$12$}
    \psfrag{14}[][]{\small$14$}
    \psfrag{16}[][]{\small$16$}
    \psfrag{18}[][]{\small$18$}
    \psfrag{20}[][]{\small$20$}
    \psfrag{100}[r][r]{\small$10^0$}
    \psfrag{101}[r][r]{\small$10^{-1}$}
    \psfrag{102}[r][r]{\small$10^{-2}$}
    \psfrag{103}[r][r]{\small$10^{-3}$}
    \psfrag{104}[r][r]{\small$10^{-4}$}
    \psfrag{105}[r][r]{\small$10^{-5}$}
    \centering
    \includegraphics[width=12cm]{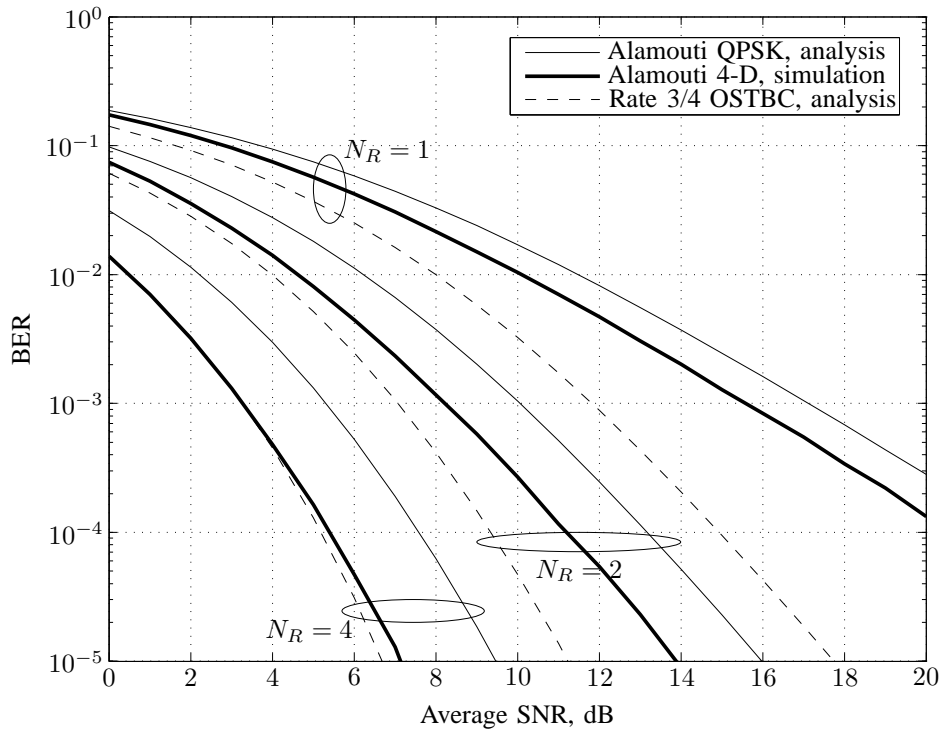}
    \caption{The BERs of three codes, Alamouti's code with
    constituent QPSK signals, the rate $3/4$ OSTBC with
    constituent QPSK signals, and Alamouti's code with
    constituent biorthogonal 4-D signals, for the Rayleigh
    fading channel.}
    \label{fig_BER_4dcmp}
\end{figure}

\begin{figure}
    \psfrag{SNR, dB}[][]{\raisebox{-20pt}{\small Average SNR, dB}}
    \psfrag{BER}[][]{\raisebox{25pt}{\small Probability of Error}}
    \psfrag{Alamouti QPSK, analysis}{\small Alamouti QPSK, analysis}
    \psfrag{Alamouti BPSK, analysis}{\small Alamouti BPSK, analysis}
    \psfrag{Alamouti 8-D, simulation}{\small Alamouti 8-D, simulation}
    \psfrag{NR = 1}{\small$N_R = 1$}
    \psfrag{NR = 2}{\small$N_R = 2$}
    \psfrag{NR = 4}{\small$N_R = 4$}
    \psfrag{0}[][]{\small$0$}
    \psfrag{2}[][]{\small$2$}
    \psfrag{4}[][]{\small$4$}
    \psfrag{6}[][]{\small$6$}
    \psfrag{8}[][]{\small$8$}
    \psfrag{10}[][]{\small$10$}
    \psfrag{12}[][]{\small$12$}
    \psfrag{14}[][]{\small$14$}
    \psfrag{16}[][]{\small$16$}
    \psfrag{18}[][]{\small$18$}
    \psfrag{20}[][]{\small$20$}
    \psfrag{100}[r][r]{\small$10^0$}
    \psfrag{101}[r][r]{\small$10^{-1}$}
    \psfrag{102}[r][r]{\small$10^{-2}$}
    \psfrag{103}[r][r]{\small$10^{-3}$}
    \psfrag{104}[r][r]{\small$10^{-4}$}
    \psfrag{105}[r][r]{\small$10^{-5}$}
    \centering
    \includegraphics[width=12cm]{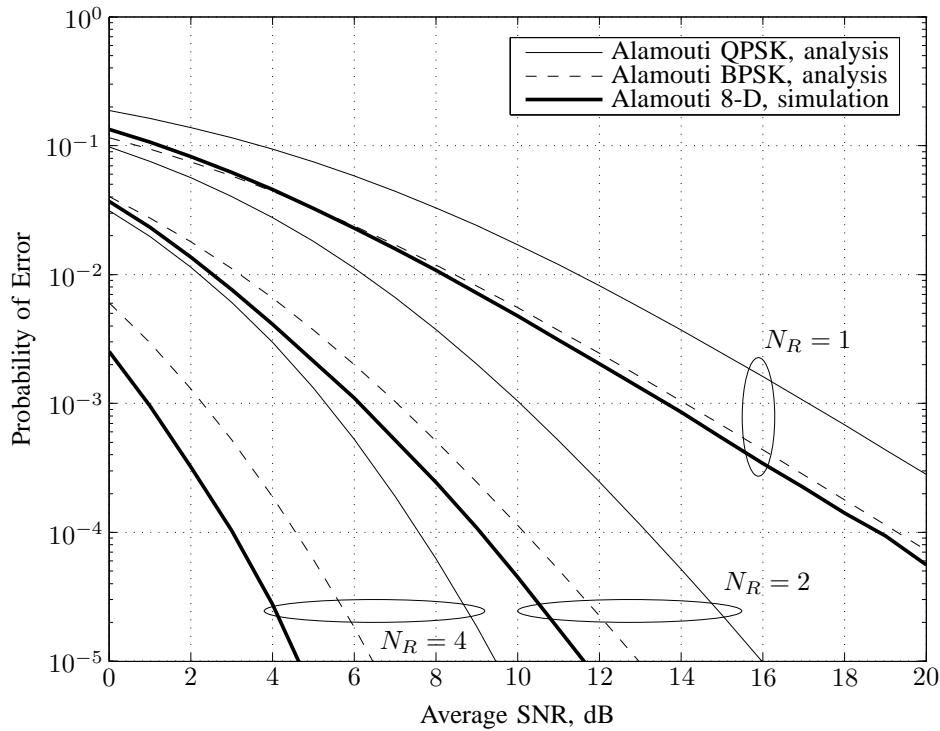}
    \caption{The BERs of three codes, Alamouti's code with
    constituent QPSK signals, Alamouti's code with constituent
    BPSK signals, and Alamouti's code with constituent
    biorthogonal 8-D signals, for the Rayleigh fading channel.}
    \label{fig_BER_8dcmp}
\end{figure}

Figs. \ref{fig_BER_4dcmp} and \ref{fig_BER_8dcmp} show the
simulated BER performances of the OSTBC designs based on spherical
codes presented in Section \ref{codedesign}. The receiver
structure is ML decoding based on the equivalent model shown in
Fig. \ref{fig_EquiModel}. In fact, the signaling for the
equivalent model is $M$-ary biorthogonal \cite{ProakisJG08}. The
simulations have been done for Rayleigh fading with $10^7$ trials.

Fig. \ref{fig_BER_4dcmp} also shows the BERs of two other
conventional schemes for comparison, Alamouti's code, and the rate
$3/4$ OSTBC \eqref{eq13}, both with constituent QPSK signals. Note
that in these codes, the constituent signals are independent (in
contrast to the new design) so that the Alamouti code has $M =
16$, and the rate $3/4$ code has $M = 64$. The BERs of these codes
are known to be equivalent to the BER of QPSK signaling in
Nakagami fading channels. The latter has been analytically
obtained in \cite[Section 5.1]{SimonMK05}. The choice of these two
codes for comparison is justified as follows. The new code based
on 4-D spherical codes uses two transmitting antennas and its rate
is $3/2$ bits per 2-D degree of freedom (DoF). However, there is
no conventional space-time code with the same rate that uses two
transmitting antennas. Nonetheless, the Alamouti code with
constituent QPSK signals uses two transmitting antennas, but its
rate is $2$ bits per 2-D DoF. The rate $3/4$ code uses four
transmitting antennas, but has the same rate as the new code (note
that the label ``rate $3/4$'' only refers to the fact that the
system transmits three symbols in four time slots).

It can be seen from Fig. \ref{fig_BER_4dcmp} that the new design
is superior to the conventional Alamouti scheme. However, the new
design is outperformed by the OSTBC \eqref{eq13}, especially for a
smaller number of antennas. Note that the new code uses half as
many transmitting antennas as the OSTBC \eqref{eq13}, which
translates into less complexity and smaller size.

Fig. \ref{fig_BER_8dcmp}, in a manner similar to Fig.
\ref{fig_BER_4dcmp}, exhibits and compares the BER performances of
three codes, including the new OSTBC design based on 8-D spherical
codes, and two conventional schemes, Alamouti's code with
constituent BPSK signals and Alamouti's code with constituent QPSK
signals. The QPSK Alamouti code here is the same as the one used
for comparison in Fig. \ref{fig_BER_4dcmp}. The BER of the BPSK
Alamouti scheme has been obtained from \eqref{eq33}. Note that all
the codes use two transmitting antennas, and their rates are
respectively $1$ bit, $1$ bit, and $2$ bits per 2-D DoF. Fig.
\ref{fig_BER_8dcmp} demonstrates the BER superiority of the new
code over the two conventional schemes, i.e., even over the BPSK
Alamouti scheme which has the same rate. The superiority is
augmented as the number of receiving antennas increases.

\begin{table}
\caption{Performance Tradeoffs in Transition From Alamouti's Code
With Constituent Biorthogonal \protect\\4-D Signals to the Code
With Biorthogonal 8-D Signals} \vspace{-20pt}
\begin{center}
    \begin{tabular}{cccccc}
        \toprule\addlinespace[1pt]
        \rule{0pt}{13pt} Time DoFs & Frequency DoFs & Power
        & SNR & Bit Rate & BER \\\addlinespace[1pt]
        \hline
        \rule{0pt}{13pt} $1$ & $2$ & $1$ & $0.5$ & $4/3$
        & $\approx 1.6 N_R^{0.33}$ \\\addlinespace[1pt]
        \hline
        \rule{0pt}{13pt} $1$ & $2$ & $2$ & $1$ & $4/3$
        & $\approx 0.50 N_R^{-1.7}$ \\\addlinespace[1pt]
        \hline
        \rule{0pt}{13pt} $2$ & $1$ & $1$ & $1$ & $2/3$
        & $\approx 0.50 N_R^{-1.7}$ \\\addlinespace[1pt]
        \hline
        \rule{0pt}{13pt} $2$ & $1$ & $2$ & $2$ & $2/3$
        & $\approx 0.11 N_R^{-3.3}$ \\
        \bottomrule
    \end{tabular}
\end{center}
\label{tab_alamcomp}
\end{table}

\begin{figure}
    \psfrag{Eb/N0, dB}[][]{\raisebox{-20pt}{\small$E_b / N_0$, dB}}
    \psfrag{Probability of Error}[][]{\raisebox{25pt}{\small Probability of Error}}
    \psfrag{SER, analytical}{\small SER, analytical}
    \psfrag{BER, analytical}{\small BER, analytical}
    \psfrag{Simulation results*****}{\small Simulation results}
    \psfrag{NR = 1}{\small$N_R = 1$}
    \psfrag{NR = 2}{\small$N_R = 2$}
    \psfrag{NR = 4}{\small$N_R = 4$}
    \psfrag{0}[][]{\small$0$}
    \psfrag{2}[][]{\small$2$}
    \psfrag{4}[][]{\small$4$}
    \psfrag{6}[][]{\small$6$}
    \psfrag{8}[][]{\small$8$}
    \psfrag{10}[][]{\small$10$}
    \psfrag{12}[][]{\small$12$}
    \psfrag{14}[][]{\small$14$}
    \psfrag{16}[][]{\small$16$}
    \psfrag{18}[][]{\small$18$}
    \psfrag{20}[][]{\small$20$}
    \psfrag{100}[r][r]{\small$10^0$}
    \psfrag{101}[r][r]{\small$10^{-1}$}
    \psfrag{102}[r][r]{\small$10^{-2}$}
    \psfrag{103}[r][r]{\small$10^{-3}$}
    \psfrag{104}[r][r]{\small$10^{-4}$}
    \psfrag{105}[r][r]{\small$10^{-5}$}
    \centering
    \vspace{100pt}
    \includegraphics[width=12cm]{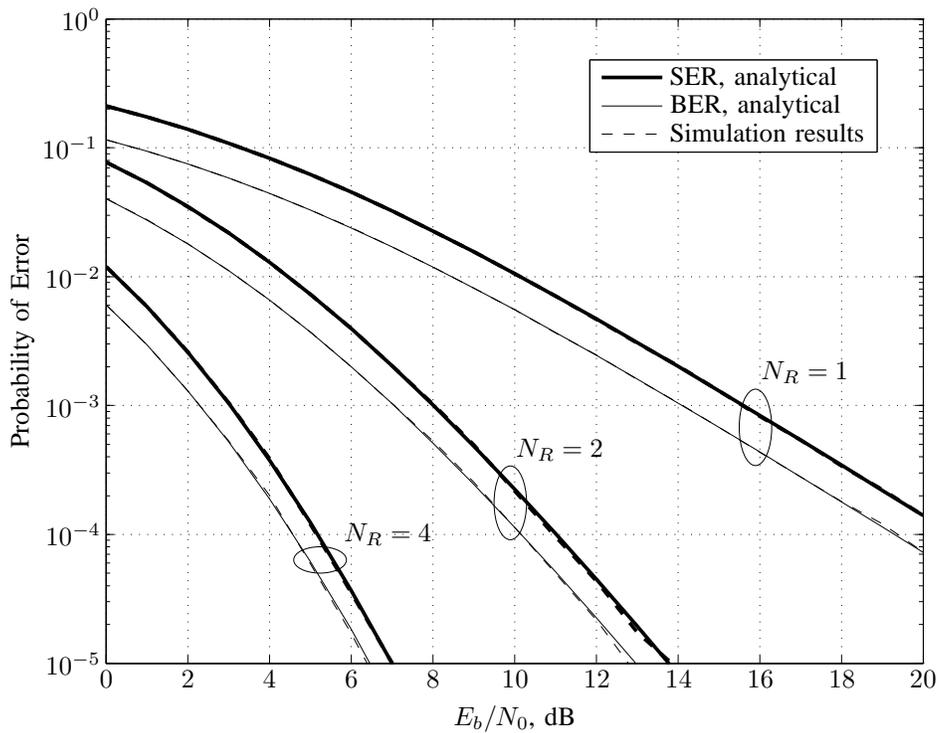}
    \caption{The SER and BER of Alamouti's code with constituent
    BPSK signals for the Rayleigh fading channel.}
    \vspace{100pt}
    \label{fig_SER_BER_AlamBPSK}
\end{figure}

It is noteworthy that the performances of the new designs based on
4-D and 8-D spherical codes in Figs. \ref{fig_BER_4dcmp} and
\ref{fig_BER_8dcmp} are not directly comparable. In fact, in
transition from the code with $n = 4$ and $M = 8$ (Fig.
\ref{fig_BER_4dcmp}) to the code with $n = 8$ and $M = 16$ (Fig.
\ref{fig_BER_8dcmp}), the number of dimensions or DoFs is doubled,
which means that twice as many time and/or frequency resources are
expended. The impact of this transition is shown in Table
\ref{tab_alamcomp} in terms of tradeoffs between the utilized time
DoFs, utilized frequency DoFs, continuous-time transmitting power,
SNR, bit rate, and BER. Four different cases of tradeoff have been
represented. In the table, value $x$ denotes that the value of the
corresponding quantity is multiplied by $x$ as a result of the
transition. The approximate changes shown for the BER are obtained
by a comparison between Figs. \ref{fig_BER_4dcmp} and
\ref{fig_BER_8dcmp} for relatively large values of SNR.

Finally, Fig. \ref{fig_SER_BER_AlamBPSK} shows the SER and BER
performances of Alamouti's code with BPSK constituent signals for
Rayleigh fading. The figure exhibits both results based on the
analysis presented in Section \ref{serberanalysis}, and simulation
results. In the analytical approach, integral \eqref{eq35} has
been evaluated numerically. Note that the simulation results are
in excellent agreement with the analytical results.

\section{Conclusion}
Based on the analysis of the distance properties of OSTBCs, an
equivalent model for MIMO communication system with OSTBCs was
proposed and a class of Euclidean codes equivalent to OSTBCs was
introduced. Examples of distance spectra, signal constellations,
and signal coordinate diagrams of Euclidean codes equivalent to
some simplest OSTBCs were given. A new asymptotic upper bound on
the SER of OSTBCs, which is based on the distance spectra of the
introduced equivalent Euclidean codes, was derived. Also, new
general design criteria for the signal constellations of optimal
OSTBCs were proposed for two asymptotic cases, (i) large SNR and
(ii) large SNR and a large number of antennas. Exploiting the
connection between OSTBCs and spherical codes, some bounds which
link the distance properties, dimensionality, and cardinality of
equal energy OSTBC signals were given. Then, two new optimal
signal constellations with cardinalities $M = 8$ and $M = 16$ for
Alamouti's code were designed as an example of using the
connection between OSTBCs and spherical codes. Finally, using the
model introduced for MIMO communication systems with OSTBCs, a
general methodology for performance analysis of OSTBCs was
formulated. As an application example of this methodology, a new
expression for the SER of Alamouti's code with BPSK signals was
derived. This result is the first example of exact SER analysis of
OSTBCs.

\bibliography{mybibfile}
\bibliographystyle{IEEEtran}
\end{document}